\documentclass[twocolumn,trackchanges]{aastex701}

\usepackage{amsmath}
\usepackage{multirow}
\usepackage{subcaption}

\usepackage{xcolor}

\begin{document}

\title{Spiral Structure Properties, Dynamics, and Evolution in MW-mass Galaxy Simulations}

\author[0000-0001-6086-9873]{J. R. Quinn}
\affiliation{Department of Physics, University of California, Merced, 5200 N. Lake Road, Merced, CA 95343, USA}
\email{jquinn4@ucmerced.edu}

\author[0000-0003-3217-5967]{S. R. Loebman}
\affiliation{Department of Physics, University of California, Merced, 5200 N. Lake Road, Merced, CA 95343, USA}
\email{sloebman@gmail.com}

\author[0000-0003-2594-8052]{K. J. Daniel}
\affiliation{Department of Astronomy \& Steward Observatory, University of Arizona,
Tucson, AZ 85721, USA}
\email{drkatedaniel@gmail.com}

\author[0000-0002-0740-1507]{L. {Beraldo e Silva}}
\affiliation{Department of Astronomy \& Steward Observatory, University of Arizona,
Tucson, AZ 85721, USA}
\affiliation{Observatório Nacional, Rio de Janeiro - RJ, 20921-400, Brasil}
\email{lberaldoesilva@on.br}

\author[0000-0003-0603-8942]{A. Wetzel}
\affiliation{Department of Physics \& Astronomy, University of California, Davis, CA 95616, USA}
\email{arwetzel@gmail.com}

\author[0000-0001-7902-0116]{V. P. Debattista}
\affiliation{Jeremiah Horrocks Institute, University of Central Lancashire, Preston PR1 2HE, UK}
\email{vpdebattista@gmail.com}

\author[0000-0002-8354-7356]{A. Arora}
\affiliation{Department of Astronomy, University of Washington, Seattle, WA 98195, USA}
\email{arora125@uw.edu}

\author[0000-0002-3353-6421]{S. Ansar}
\affiliation{Institute for Computational Cosmology, Department of Physics, Durham University, South Road, Durham DH1 3LE, UK}
\affiliation{Centre for Extragalactic Astronomy, Department of Physics, Durham University, South Road, Durham DH1 3LE, UK}
\email{sioree.ansar@durham.ac.uk}

\author{F. McCluskey}
\affiliation{Department of Physics \& Astronomy, University of California, Davis, CA 95616, USA}
\email{fmccluskey@ucdavis.edu}

\author[0009-0009-1128-5746]{D. Masoumi}
\affiliation{Department of Physics, University of California, Merced, 5200 N. Lake Road, Merced, CA 95343, USA}
\email{dmasoumi@ucmerced.edu}

\author[0000-0001-6380-010X]{J. Bailin}
\affiliation{Department of Physics \& Astronomy, University of Alabama, Box 870324, Tuscaloosa, AL 35487-0324, USA}
\email{jbailin@ua.edu}

\begin{abstract}

The structure of spiral galaxies is essential to understanding the dynamics and evolution of disk galaxies; however, the precise nature of spiral arms remains uncertain. 
Two challenges in understanding the mechanisms driving spirals are how galactic environment impacts spiral morphology and how they evolve over time.
We present a catalog characterizing the properties, dynamics, and evolution of $m=2$ spiral structure in 10 Milky Way-mass galaxies from the FIRE-2 cosmological zoom-in simulations.
Consistent with previous literature, we find that FIRE-2 spirals are transient, recurring features simultaneously present in the disk at varying pattern speeds ($\Omega_p$) that broadly decrease with radius. 
These spirals persist on Gyr timescales (mean duration 1.90~Gyr), but fluctuate in amplitude on timescales of hundreds of Myr. 
Tidal interactions and bar episodes impact the resulting $m=2$ spiral structure; strong satellite interactions generally produce shorter-lived, stronger spirals with larger radial extent, and bars can increase $\Omega_p$. 
Galactic environment influences spiral structure; kinematically colder disks can support longer-lived, stronger spirals. 
The properties of identified spirals in FIRE-2 vary widely in radial extent (0.3-10.8~kpc), duration (1.00-6.00~Gyr), and amplitudes ($a_{2,\text{max}}$=0.018-0.192). 
We find the presence of spirals in all age populations, suggesting these are density wave-driven features.
This work represents the first time that spiral structure has been cataloged in this manner in cosmological simulations; the catalog can be leveraged with current and forthcoming observational surveys, enabling systematic comparisons to further our understanding of galaxy evolution.

\end{abstract}


\section{Introduction} \label{sec:intro}
Spiral structure is seen ubiquitously among disk galaxies in our local Universe and exhibits remarkable morphological variation across diverse galactic environments \citep{baillard2011,lintott2011,willett2013}.
Characterized by curved arms in a flattened disk of stars and gas, these features can appear well-defined and continuous over a large angular extent (grand design), fragmented and patchy (flocculent), or a combination of both (multi-armed) \citep{elmegreen&elmegreen1982,kendall&kennicutt2011}. 
They are observed in both blue gas-rich and red quiescent galaxies \citep{bamford2009,masters2010,hart2016,tojeiro2013,cui2024} and exist at both low and high redshift \citep{elmegreen&elmegreen2014,costantin2023,huang2023,kuhn2023,wu2023}.

Spiral arms serve as predominant regions of star formation \citep{roberts1969,yu2022,peltonen2024,williams2024} and are fundamentally connected to the dynamic and evolutionary processes of disk galaxies by redistributing material throughout the galactic disk \citep{sellwood&binney2002,grand2012b,roskar2012}.
For a comprehensive review on these topics, see \cite{dobbs&baba2014} and \cite{sellwood&masters2022}.
Despite the significance spiral arms play in galaxy evolution, the dominant physical processes that drive their formation and persistence remains an open question in astrophysics.

Many theories have been proposed to explain the nature of spiral arms, such as density waves \citep{lin&shu1964, bertin&lin1996, bertin1989}, swing amplification \citep{goldreich&lynden-bell1965, julian&toomre1966}, bar-driven responses \citep[e.g.,][]{kormendy&norman1979, tremaine&weinberg1984, garma-oehmichen2021}, tidally-driven responses \citep[e.g.,][]{salo&laurikainen2000, dobbs2010,purcell2011,pettitt2017}, and dynamical mechanisms that maintain self-excited spirals \citep{athanassoula2012,d'onghia2013,sellwood&carlberg2014,sellwood&carlberg2019,sellwood&carlberg2021,sellwood&carlberg2022}. 
Numerical simulations provide a valuable tool for testing the plausibility and significance of each of these theories in driving the formation and persistence of spirals. 
Early N-body simulations showed that spiral arms are short-lived, transient, recurring structures \citep{sellwood&carlberg1984,carlberg&freedman1985}.
Simulations to date continue to demonstrate the transient nature of spiral structure; spiral arms continuously break apart and reform and have pattern speeds that follow the galaxy's differential rotation \citep{wada2011,grand2012a,roskar2012,baba2013}. 
High resolution simulations have also suggested that spiral arms can live for much longer---on the scale of several billion years---than previously thought, through a self-regulating mechanism that maintains and regenerates spirals \citep{fujii2011,d'onghia2013}.

Such numerical studies are particularly important for understanding the formation and evolution of our Milky Way (MW).
Previous studies have worked to understand the MW's spiral morphology \citep{georgelin&georgelin1976,fux1997,fux1999,drimmel2000,benjamin2005,hou2009,hou&han2014}, which has become increasingly achievable with the advent of better observational data from Gaia \citep{gaia2018}, the Bar and Spiral Structure Legacy (BeSSeL) Survey \citep{brunthaler2011}, and the VLBI Exploration of Radio Astrometry (VERA) project \citep{vera-collab2020}. 
The MW is commonly depicted as having four main spiral arms \citep{vallee2017,reid2019,minniti2021}; however, it is difficult to determine the true spiral morphology due to our location in the Galactic mid-plane. 
Despite this challenge, data from BeSSeL and VERA supports the view of the MW as a four-armed spiral with several additional arm segments and spurs \citep{reid2019}. 
However, using VLBI observations and Gaia DR3, \cite{xu2023} contends that the MW has two primary spiral arms in the inner disk and bifurcates further out. 
Moreover, different stellar populations reveal different spiral structure. Young stars show patchy, multi-armed features \citep{poggio2021, drimmel2025}, whereas older stars display a more grand design-like two-armed spiral \citep{churchwell2009, khanna2025}.

Studies have tried to understand the origin of the MW's spirals; \cite{eilers2020} used Gaia DR2 data to reveal kinematic signatures of dynamical spiral arms, and \cite{baba2018}, \cite{hunt2018}, and \cite{sellwood2019} find evidence that the MW's spiral arms are transient using Gaia DR1 and DR2 data.
While advancements in observational capabilities have enabled a more precise description of the MW's spiral structure, there remains much debate about the true morphology of the MW and the underlying dynamical processes driving its evolution.

Considerable advancements in computing capabilities have led to a new generation of hydrodynamic simulations that model the evolution of galaxies in a cosmological context. 
Large volume simulations such as Illustris \citep{vogelsberger2014}, IllustrisTNG \citep{weinberger2017, pillepich2018}, and EAGLE \citep{crain2015, schaye2015} simulate a large number of realistic galaxies but have coarser particle mass resolution.
Zoom-in MW-mass simulations such as Auriga \citep{grand2017, grand2024} and the FIRE-2 \textit{Latte} galaxies \citep{wetzel2016,hopkins2018,wetzel2023} contain fewer galaxies but are able to resolve detailed dynamics of the disk in a fully cosmological context.

The primary aim of this paper is to describe the properties, dynamics, and time-evolution of spiral structure in high resolution, hydrodynamic, cosmological MW-mass FIRE-2 galaxy simulations.
A secondary aim is to understand how varying galactic environments and evolutionary histories affects the resulting spiral structure.
Our ultimate goal is to understand the physics that drives the formation and evolution of spiral structure in these galaxies.
This first catalog paper provides a systematic basis for future works that will study the physical drivers of spiral structure.
That is, by utilizing simulations that model galaxies more realistically than ever before, we enable investigations of which mechanisms are most likely driving spiral arms to form and persist in these galaxies.

\S\ref{sec:methods} describes the simulations used in this study and discusses the methodology for identifying and analyzing spiral structure. 
\S\ref{sec:results} presents the results of our analysis. 
In \S\ref{sec:disc}, we discuss how our findings fit into the current understanding of spiral structure in simulations.
Finally, we summarize our results in \S\ref{sec:conc}.

\section{Methods}\label{sec:methods}
\subsection{FIRE-2 Simulations} \label{ssec:sims}

This paper analyzes cosmological hydrodynamic zoom-in simulations of Milky Way-mass galaxies from the Feedback in Realistic Environments (FIRE) project. 
The simulations are run using the FIRE-2 physics model \citep{hopkins2018} and use the GIZMO \citep{hopkins2015} gravity+hydrodynamics code with the Mesh-free Finite Mass (MFM) mode. 
MFM is a Lagrangian method that maintains conservation of mass, energy, and (angular) momentum while providing adaptive spatial resolution. 
FIRE-2 simulations implement metallicity-dependent radiative heating and cooling which span a temperature range of 10 - 10$^{10}$ K. 
The simulations include radiative heating from a spatially uniform, redshift-dependent ionizing UV background \citep{faucher-giguere2009}.

Star formation occurs in gas that is locally self-gravitating, Jeans-unstable, dense (n $>$ 1000 cm$^3$), and molecular \citep[following][]{krumholz&gnedin2011}.
Once a gas cell becomes eligible for star formation, it converts to a star particle on a local free-fall time representing a stellar population with mass and elemental abundance inherited from the gas cell with a \cite{kroupa2001} initial mass function.

FIRE-2 models time-resolved stellar feedback processes, including continuous mass loss from stellar winds, core-collapse and white-dwarf (Ia) supernovae, radiation pressure, photoionization, and photo-electric heating. 
Star particles return mass, metals, momentum, and energy back into the ISM at a rate following the \texttt{STARBURST99} stellar evolution models \citep{leitherer1999}.

We generated cosmological zoom-in initial conditions at $z\approx 99$ using MUSIC \citep{hahn&abel2011}. 
All simulation zoom-in regions are embedded within a cosmological box with side length 70.4 - 172 Mpc. 
The simulations assume a flat $\Lambda$CDM cosmology with parameters that are broadly consistent with the \cite{planck-collaboration2020}: $h = 0.68-0.71$, $\Omega_\Lambda = 0.69-0.734$, $\Omega_m = 0.266-0.31$, $\Omega_b = 0.0455-0.048$, $\sigma_8 = 0.801-0.82$, and $n_s = 0.961-0.97$.

We utilize 10 MW/M31 mass galaxies, consisting of 5 isolated galaxies from the \textit{Latte} suite introduced in \cite{wetzel2016} and 5 galaxies from the \textit{Elvis on FIRE} suite of Local Group-like (LG-like) pairs of galaxies \citep{garrison-kimmel2019a, garrison-kimmel2019b}. 
The isolated galaxies have an initial baryon particle mass of approximately 7100 M$_\odot$. 
However at present day most particles have a mass of approximately 5000 M$_\odot$ due to stellar mass loss. 
Romeo and Juliet have initial particle masses of 3500 M$_\odot$, while Romulus and Remus and Thelma and Louise have initial particle masses of 4000 M$_\odot$. 
Stellar and dark-matter particles have fixed gravitational force softenings (comoving at z $>$ 9 and physical at z $<$ 9) with a Plummer equivalent of $\epsilon_{\text{star}}=$4 pc and $\epsilon_{\text{dm}}=$ 40 pc. 
Gas cells have adaptive softening that matches the hydrodynamic kernel smoothing and reaches a minimum softening length of 1 pc, with a typical of approx. 20 pc in the cold ISM.

All of the analysis described in this paper has been applied to all 10 FIRE-2 galaxies. 
We decide to run the analysis on these 10 galaxies from within the \textit{Latte} and \textit{Elvis on FIRE} suite as these galaxies all transition to a rate of steady star formation and become rotationally supported thin disks for a sufficient length of time for us to run our analysis over. 
Table~\ref{tab:properties} shows properties of the FIRE-2 galaxies analyzed in this paper at present day. 
Among these 10 simulations, we have chosen two to highlight throughout this paper, m12f and m12m.
These two galaxies represent examples of galaxies with very different merger histories and galactic environments and show bracketing extremes of Milky Way-mass galaxy evolution near more recent times.
One galaxy, m12f, undergoes more active satellite interaction, while m12m displays a more quiescent history at low $z$ \citep{sanderson2020, arora2022, horta2023}.
The two different conditions under which these galaxies evolve allows us to investigate the physical mechanisms that drive spiral structure to form and persist in FIRE-2 simulations.

A snapshot of each galaxy is saved approximately every 25 Myr. 
During the last 100~Myr, snapshots are saved every $\sim2$~Myr.
At each snapshot, we reorient the galaxy such that the angular momentum vector of all stars within 10 kpc of the galaxy's center is parallel to the z-axis. 
This alignment process is consistently applicable once the galaxy's disk has stabilized \citep{gurvich2023}. 
We align the galaxies such that they rotate in the counterclockwise sense. 
We run our analysis using data from $\sim6.8$ Gyr to present day as most of our galaxies form a thick disk around $\sim8$ Gyr ago \citep{mccluskey2023} and transition from bursty to steady star formation rate around 2~Gyr afterwards \citep{yu2021}.
See \cite{mccluskey2023} for a more in-depth description of these systems and when they become rotationally supported thin disks.

\begin{deluxetable}{cccccc}
\tablecaption{Properties of FIRE-2 galaxies at $z=0$. \label{tab:properties}}
\tablehead{
   \colhead{Simulation} & \colhead{$M_{90,*}$} & \colhead{$M_{90,\text{gas}}$} & \colhead{$R_{90}$} & \colhead{$t_{\text{B-S}}$} & \colhead{$t_{\text{onset}}$} \\
     & [$10^{10}M_{\odot}$] & [$10^{10}M_{\odot}$] & [kpc] & [Gyr] & [Gyr]
}
\startdata
    m12i$^1$ & $6.3$ & $1.0$ & 10.0 & 3.14 & 6.65 \\
    m12f$^2$ & $7.9$ & $1.5$ & 13.3 & 5.01 & 7.42 \\
    m12m$^3$ & $11.0$ & $1.7$ & 12.5 & 3.81 & 9.21 \\
    m12b$^4$ & $8.5$ & $1.2$ & 10.9 & 6.32 & 7.42 \\
    m12c$^4$ & $5.8$ & $1.1$ & 10.4 & 3.70 & 6.49 \\
    Romeo$^4$ & $6.6$ & $1.2$ & 13.3 & 6.52 & 11.0 \\
    Juliet$^4$ & $3.8$ & $0.7$ & 9.6 & 4.40 & 4.35 \\
    Romulus$^5$ & $9.1$ & $1.9$ & 14.2 & 4.90 & 7.42 \\
    Remus$^5$ & $4.6$ & $1.4$ & 12.3 & 5.88 & 7.93 \\
    Thelma$^4$ & $7.1$ & $1.6$ & 12.4 & 2.57 & 4.35 
\enddata
\tablecomments{Columns: (1) Name of the simulation and its reference; simulation names starting with `m12' indicate an isolated galaxy from the \textit{Latte} suite, while those that do not name galaxies in Local Group-like pairs from \textit{Elvis on FIRE}. 
(2) $R_{90}$: The radius enclosing 90\% of the stellar mass within the central 20 kpc in spherical coordinates. 
(3) $M_{90,*}$: The total stellar mass enclosed at $R_{90}$. 
(4) $M_{90,\text{gas}}$: The total gas mass enclosed at $R_{90}$. 
(5) $t_{\text{B-S}}$: The lookback time when each galaxy transitions from bursty to steady star formation from \cite{yu2021}. 
(6) $t_{\text{onset}}$: The lookback time for onset of disk formation, defined as the time when $(\frac{v_{\phi}}{\sigma_{\text{tot}}})_{\text{form}}>1$ from \cite{mccluskey2023}.
References are as follows: 1-\cite{wetzel2016}; 2-\cite{garrison-kimmel2017}; 3-\cite{hopkins2018}; 4-\cite{garrison-kimmel2019a}; 5-\cite{garrison-kimmel2019b}}
\end{deluxetable}
\begin{figure}
\includegraphics[width=\linewidth]{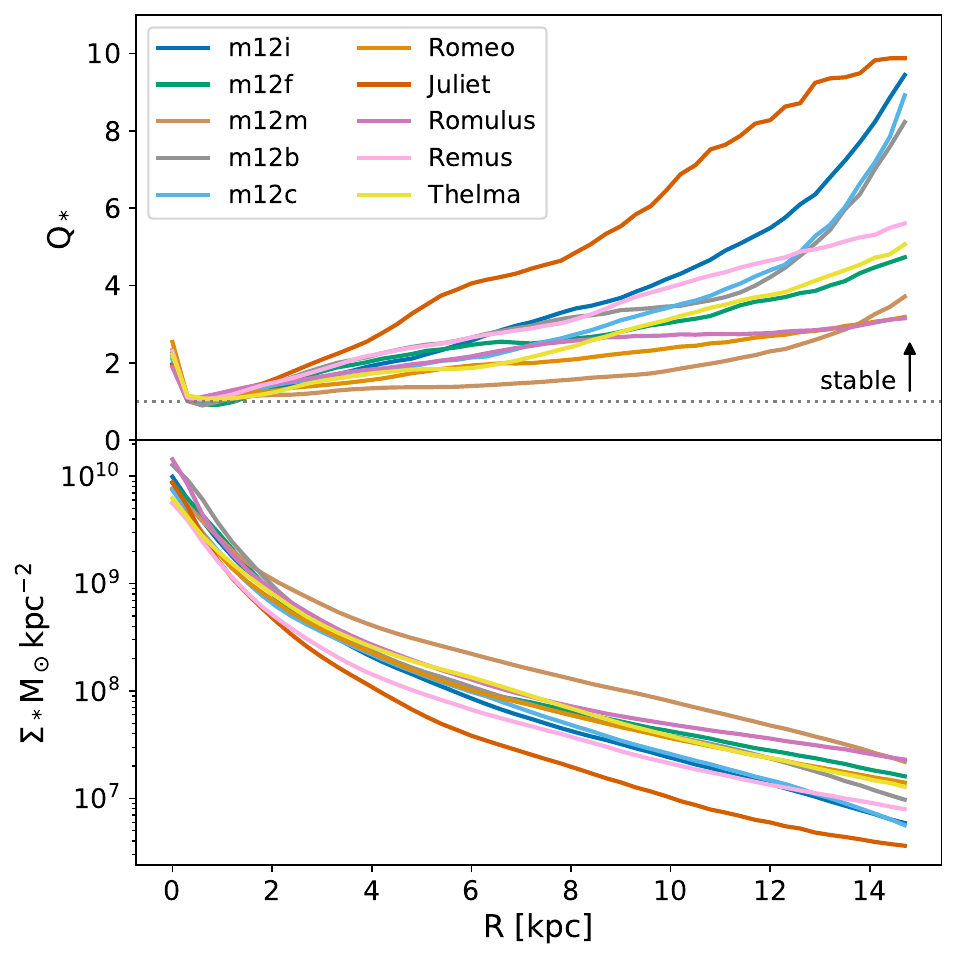}
\caption{Top: Measure of disk stability (Stellar Toomre $Q_*$, Eq.~\ref{eq:1}) for each simulation at $z=0$, as a function of radius. 
The stability criterion ($Q>1$) is indicated by the horizontal dotted line. 
Bottom: Stellar surface density for each simulation at present day, as a function of radius.
The FIRE-2 galaxies we analyze are stable ($Q>1$) beyond the central few kpc, which allows for the presence of recurring spirals in our simulations.
\label{fig:ToomreQ}}
\end{figure}

The top panel of Fig.~\ref{fig:ToomreQ} shows the stellar Toomre Q $Q_*$ for all FIRE-2 galaxies we analyze as a function of cylindrical radius.
The Toomre Q value was calculated by
\begin{equation} \label{eq:1}
    Q_*=\sigma_R \kappa/3.36\Sigma_*G
\end{equation}
where $\sigma_R$ is the radial velocity dispersion of stars, $\kappa$ is the epicyclic frequency, $\Sigma_*$ is the surface density of stars, and $G$ is the gravitational constant.
We obtain $\kappa$ using $\kappa^2 = R (d\Omega^2/dR) + 4\Omega^2$ \citep{binney&tremaine2008}, assuming a nearly flat rotation curve.
The bottom panel shows the stellar surface density $\Sigma_*$ for the sample.
Beyond the central few kpc, our disks' stellar density distribution roughly follow a single exponential profile.
The horizontal gray dotted line indicates $Q_*=1$ and is the limit for a disk to be marginally stable to perturbations, thus allowing for longer-lived, recurring transient spirals to be present \citep{toomre1964}. 
All galaxies analyzed meet the Toomre Q stability criterion for recurring spirals to occur throughout the time that we perform our analysis over. 

\subsection{Windowed Discrete Fourier Transform Analysis} \label{ssec:WDFT}

We use a windowed discrete Fourier Transform (WDFT) spectral analysis technique to analyze the evolution of spiral amplitudes, introduced in \cite{sellwood&athanassoula1986}. 
WDFT is used to quantify the power of spiral structure and identify their spatial locations and pattern speeds. 
This is used to identify dominant spiral amplitudes by analyzing how the stellar mass distribution at each snapshot evolves over a period of time.

We begin our analysis by radially binning, using 2D cylindrical radius, all stars into $0.3$~kpc annuli at a given snapshot. 
To minimize the selection cuts we make on our particles, we do not make any cut vertically in $z$.
This means that stars in the halo and beyond the galaxy are included in the analysis.
However, we tested this analysis using various cuts in $z$ and found no appreciable difference in the results.
In each annulus, we expand the azimuthal dependence of the normalized stellar mass distribution in a Fourier series as
\begin{equation} \label{eq:1.5}
    \mu(R,\phi) = \sum_{m=0}^{\infty}c_m(R)e^{-im\phi},
\end{equation}
where the summation is over pattern multiplicity $m$ and $\phi$ is the phase, ranging $0,...,2\pi$.
The Fourier coefficients for stars in each annulus are calculated as
\begin{equation} \label{eq:2}
    c_m(R) = \frac{1}{M(R)}\sum_{p=1}^{N} m_pe^{im\phi_p}
\end{equation}
where the summation is over all stellar particles in the annulus, $M(R)$ is the total mass of all particles in the annulus, $m_p$ is the mass of the particle, $m$ is the pattern multiplicity, and $\phi_p$ is the azimuth of the particle. 
We exclude particles in the inner regions of each simulation based on the bar length at maximum strength \citep[values from Table 3][]{ansar2025}. 

\begin{figure*}
\begin{center}
\includegraphics[width=0.9\linewidth]{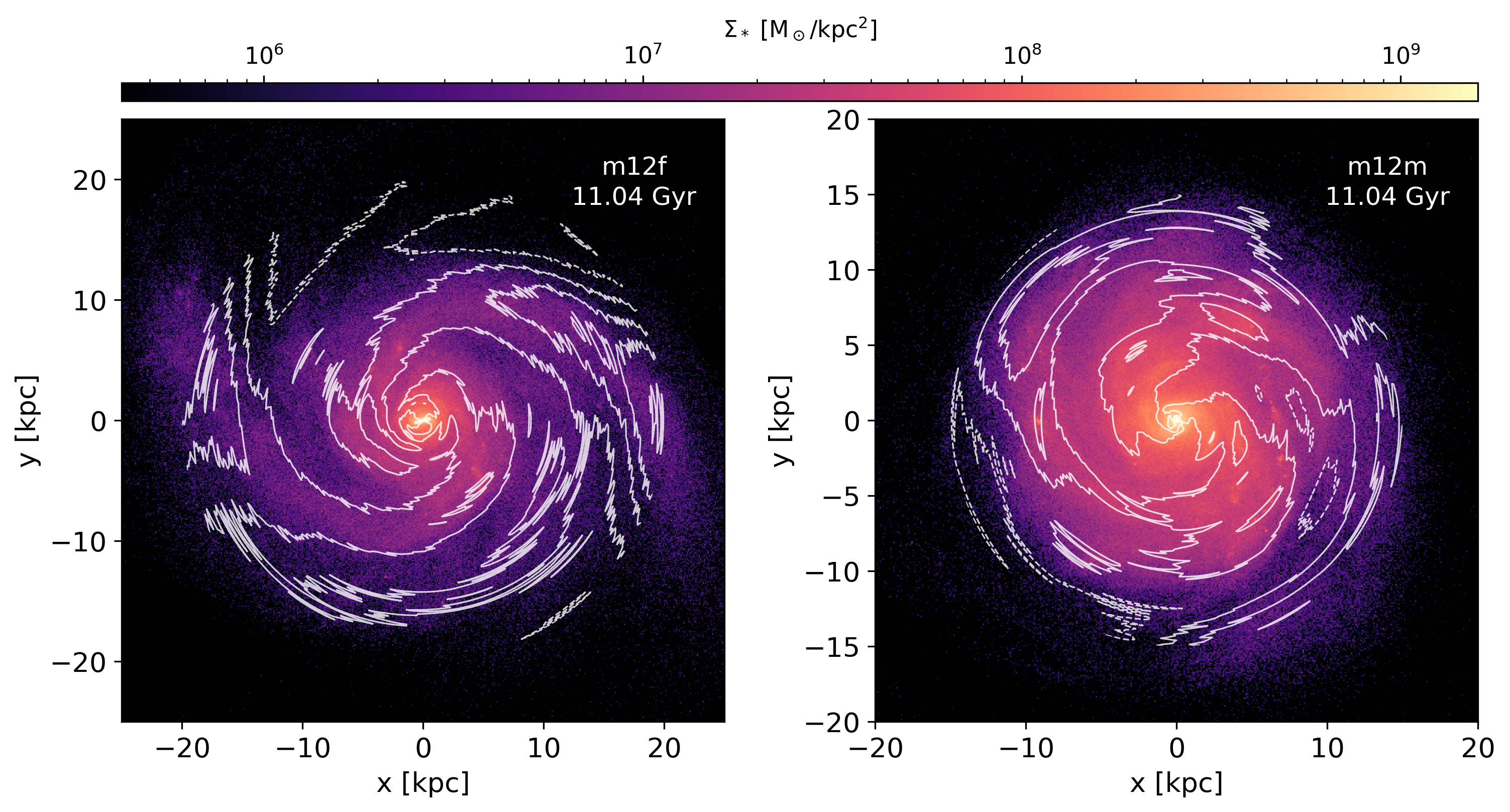}
\caption{Density contours reconstructed from the $m=1-5$ Fourier coefficients of stars $<1$~Gyr (contours) over-plotted on the stellar mass distribution of stars $<1$~Gyr (2-D histogram) for \textit{Latte} simulations m12f (left panel) and m12m (right panel).
Note the physical scale is different in each galaxy.
The solid and dashed lines show the 25, 15, and 5 percent over and under-densities, respectively.
The galaxy m12f is shown $\sim200$~Myr after a pericentric passage with a satellite with a 8:1 mass ratio and displays spiral structure that winds more tightly around the galaxy with less azimuthal spread as compared with m12m, shown after having evolved in relative isolation.
The contours of both galaxies display clear spiral structure, although their spirals look vastly different from each other and may be generated through different physical mechanisms (see \S~\ref{ssec:star_forming_gas} for further discussion).
\label{fig:contours}}
\end{center}
\end{figure*}

Using Eqs.~\ref{eq:1.5} and \ref{eq:2} we reconstruct and plot density contours of the $m=1-5$ Fourier coefficients, shown in Fig. \ref{fig:contours}.
The left panel shows m12f $\sim200$~Myr after a pericentric passage with a satellite galaxy, and the right panel shows m12m at the same time, with no recent significant external perturbations to its disk.
For better visualization of the spiral structure in Fig.~\ref{fig:contours}, we decrease the width per annulus to $0.1$~kpc and select stars $<1$~Gyr for generating the coefficients.
The stellar surface density distribution under-plotted is shown for stars $<1$~Gyr to match the contours.
While spiral structure is visually most apparent in young stars, we are able to faithfully identify the evolution dominant spiral modes in all age populations (see \S~\ref{ssec:star_forming_gas} for discussion on using different age populations).
Though the more complex features are visible with the $m=1-5$ coefficients, we see the $m=2$ multiplicity, which physically corresponds to symmetric 2-armed spirals, dominates the structure in both galaxies.
While the spirals in these galaxies are likely generated by different physical mechanisms (tidal interaction vs isolated evolution), the Fourier coefficients successfully capture the spiral structure in both systems.

\begin{figure}
\includegraphics[width=\linewidth]{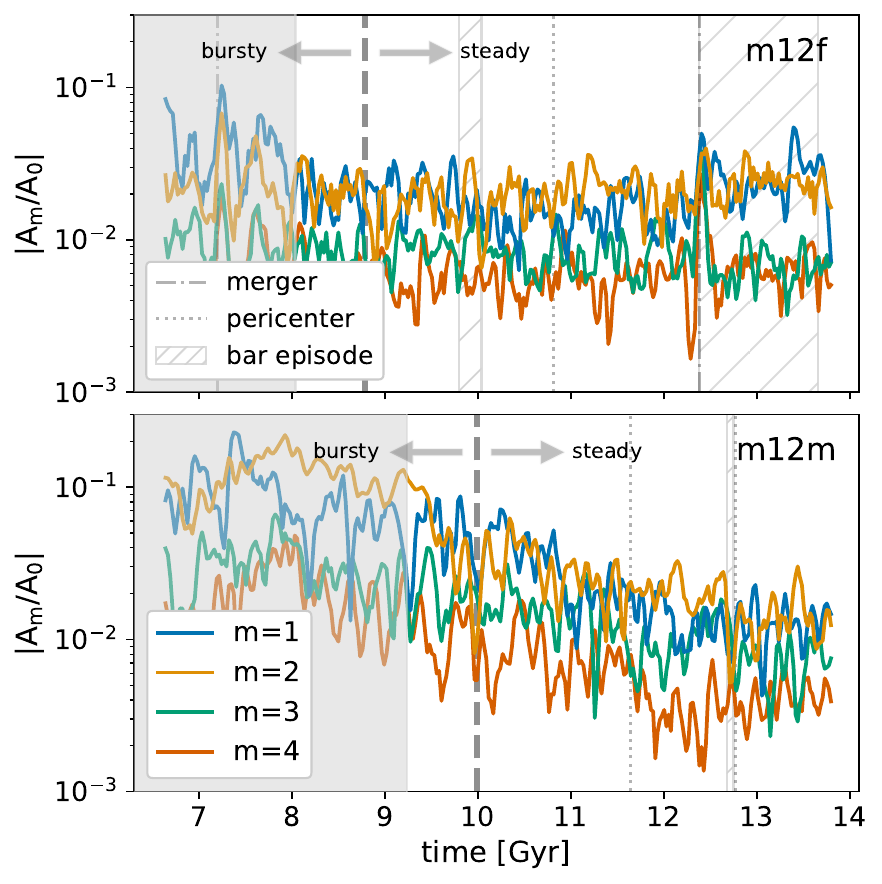}
\caption{Global Fourier amplitudes for the m=1-4 multiplicities shown across time for galaxies m12f (top panel) and m12m (bottom panel).
The thick gray vertical dashed line indicates when the galaxy transitions from bursty to steady star formation rate.
The grayed out region shows the time period that we do not analyze the galaxy.
We indicate when the galaxy undergoes merger events (dash-dot line), pericentric passages (dotted line), and bar episodes (hatched rectangle).
The $m=1$ (disk lopsidedness) and m=2 (spiral arms) amplitudes are generally dominant across time.
The $m=2$ multiplicity can be strongly impacted by bar episodes, as can be seen in m12f at late times.
All m-multiplicities are influenced by strong satellite interactions, as can be clearly seen in m12f.
\label{fig:global_amplitudes}}
\end{figure}

Fig. \ref{fig:global_amplitudes} shows the $m=1-4$ global Fourier amplitudes for galaxies m12f (top panel) and m12m (bottom panel) across time, calculated with Eq.~\ref{eq:2} using all stars and normalized by $A_0$.
We exclude the inner regions corresponding to the bar using values from \cite{ansar2025} and go out to $R_{90}$.
Each data point plotted is smoothed by applying a moving average over three snapshots for visual clarity. 
The vertical thick gray dashed line indicates the time that the galaxy transitions from having a bursty star formation rate to a steady, nearly constant star formation rate \citep{yu2021}. 
This transition also coincides with the time that these galaxies started forming their thin disk structure \citep{yu2023}.
All FIRE-2 galaxies we analyze transition during the time range between $6.8$~Gyr and $z=0$. 
We display this range of time in all figures throughout this paper, but we only consider the results using the WDFT technique (\S\ref{sec:results}) during times near and after the galaxy transitions to steady star formation.
The period of time we do not analyze is grayed out in all figures within the paper.
We also indicate times when the galaxy undergoes either a pericentric passage or merger event with a satellite galaxy and when bar episodes occur \citep{ansar2025}.

We see the dominance of $m=2$ (two-fold symmetry) and $m=1$ (lopsidedness) Fourier amplitudes across these galaxies.
The analysis in this paper focuses on the $m=2$ multiplicity, which we can attribute to two-armed spiral structure.
We focus on the $m=2$ multiplicity both because it is more dominant than other multiplicities that correspond to spiral features and because two-armed spirals are most commonly seen in observations of grand design spirals \citep{elmegreen&elmegreen1982, hart2016}.
The $m=1$ multiplicity, while not insignificant, likely corresponds to lopsidedness and asymmetry present across the stellar distribution in the galactic disk and will be considered in future work.
The $m=3-4$ multiplicities contribute to the structure of the disk, but are less dominant than the $m=2$ multiplicity for our galaxies.

We can see the short-lived impact of perturbers in Fig.~\ref{fig:global_amplitudes} from tidal interactions and bar episodes, particularly in m12f. 
Two notable peaks present in the top panel of Fig.~\ref{fig:global_amplitudes} correspond to two merger events at $7.2$~Gyr and $12.4$~Gyr.
The first merger event had a 7:1 mass ratio at time of infall \citep{arora2022}. 
The second merger’s mass ratio was 9:1 at time of infall \citep{arora2022}, and this merger triggered a burst of star formation \citep{yu2021} which is likely the reason for the increase in strength of the $m=2$ multiplicity from the time of the merger to present day.
Generally, all multiplicities are enhanced during strong satellite interactions, whereas we see an enhancement primarily in the $m=2$ multiplicity during bar episodes.

The bottom panel of Fig.~\ref{fig:global_amplitudes} shows the evolution of global Fourier amplitudes for m12m.
Broadly speaking, this galaxy does not experience substantial external impact from satellite interactions over the time of analysis.
While this galaxy does experience two pericentric passages, at $11.64$ and $12.77$~Gyr, these tidal interactions are not particularly impactful.
Their mass ratios at the time of pericenter are quite low, 129:1 and 77:1 respectively \citep{ansar2025}.
It is interesting that the $m=1$ and $m=2$ Fourier amplitudes are significantly more dominant while this galaxy is undergoing bursty star formation. 
This may be due to increased star formation along gaseous spiral arms, leading to an increase in overall stellar mass density in those regions.
This enhancement in the $m=1$ and $m=2$ multiplicities decreases in strength as the galaxy transitions to steady star formation rate and maintains very low global amplitudes throughout the rest of its evolution.
The evolution of this galaxy at late times is further explored in \cite{debattista2019}. 

Our global Fourier amplitudes are broadly $<0.1$ over the time range shown, with lower amplitudes corresponding to weaker structure.
We note that amplitudes in Fig.~\ref{fig:global_amplitudes} are calculated across the entire galactic disk, which lowers the overall amplitude of each multiplicity present due to phase differences across radius.
We show the evolution of these global amplitudes to provide a sense of which multiplicity is generally dominant across the whole disk.
The actual properties of spiral episodes in each galaxy are computed as a function of radius, and the range of amplitudes we find are discussed in \S~\ref{ssec:a2max}.
See Appendix \ref{app:global_amplitude} for plots of the global Fourier amplitudes of all galaxies.

Fourier coefficients are calculated for every snapshot in a given time interval, or time baseline.
This is calculated as $S\Delta t$, where $S$ is the total number of snapshots used in the time baseline, and $\Delta t$ is the time between snapshots.
In our analysis, $S=60$ and the time resolution of our simulation is $\Delta t\sim25$~Myr, giving a time baseline of $S\Delta t=1.5$~Gyr.
We then compute the Fourier Transform using these coefficients as
\begin{equation} \label{eq:3}
    C_{m,k}(R) = \sum_{j=0}^{S-1} c_m(R,t_j)w_je^{-2\pi ijk/S},
\end{equation}
where $k=0,1,...S-1$, the summation is over all snapshots in the time baseline, and $w_j$ is the Gaussian window function defined as
\begin{equation} \label{eq:4}
    w(j) = e^{-(j-S/2)^2/(S/4)^2}.
\end{equation}
This windowing function is used in \cite{roskar2012} and \cite{khachaturyants2022a} and is applied to minimize spectral leakage in cases when the data sampled does not produce a continuous spectrum. 
This analysis was tested using both the Gaussian window given by eq. \ref{eq:4} and a Hanning window, and we found no appreciable difference in the results. 
The discrete frequencies for which the Fourier Transform is computed are given by
\begin{equation} \label{eq:5}
    \Omega_k = \frac{2\pi}{m}\frac{k}{S\Delta t},\ \ \ k=0,1,...\frac{S}{2}.
\end{equation}
The resulting power spectrum is then calculated as
\begin{equation} \label{eq:6}
    P(R,\Omega_k) = \frac{1}{W}\left[|C_{m,k}(R)|^2 + |C_{m,S-k}(R)|^2\right],
\end{equation}
where $k=1,2,...S/2-1$, and
\begin{equation} \label{eq:7}
    W = S\sum_{j=0}^{S-1}w_j^2
\end{equation}
is the normalization factor used to take into account the windowing function. 
Summing over all radii, we find that the total power is
\begin{equation} \label{eq:total_power}
    \text{total power}(\Omega_k) = \sum_i P(R_i,\Omega_k).
\end{equation}
We find that a time baseline of $1.5~$Gyr produces a satisfactory power spectrum to analyze spiral structure.
Using too short of a time baseline will reduce spectral resolution, and we may be unable to distinguish the dominant spiral patterns present.
Using too long of a time baseline may fail to capture transient spiral patterns which evolve over the period of time.

Using Eq.~\ref{eq:5}, we obtain our frequency resolution as $\Delta\Omega\sim 2.1$~km s$^{-1}$ kpc$^{-1}$ and Nyquist frequency of $\Omega_{S/2}\sim 60$ km s$^{-1}$ kpc$^{-1}$. 
The Nyquist frequency sets the upper limit of the frequency that can be resolved.
Frequencies above the Nyquist limit are not adequately sampled and may result in aliasing, or artifacts in the resulting power spectrum.
We expect pattern speeds of spirals in the disk to exist at frequencies below the Nyquist frequency.

\begin{figure*}
\begin{center}
\includegraphics[width=0.9\linewidth]{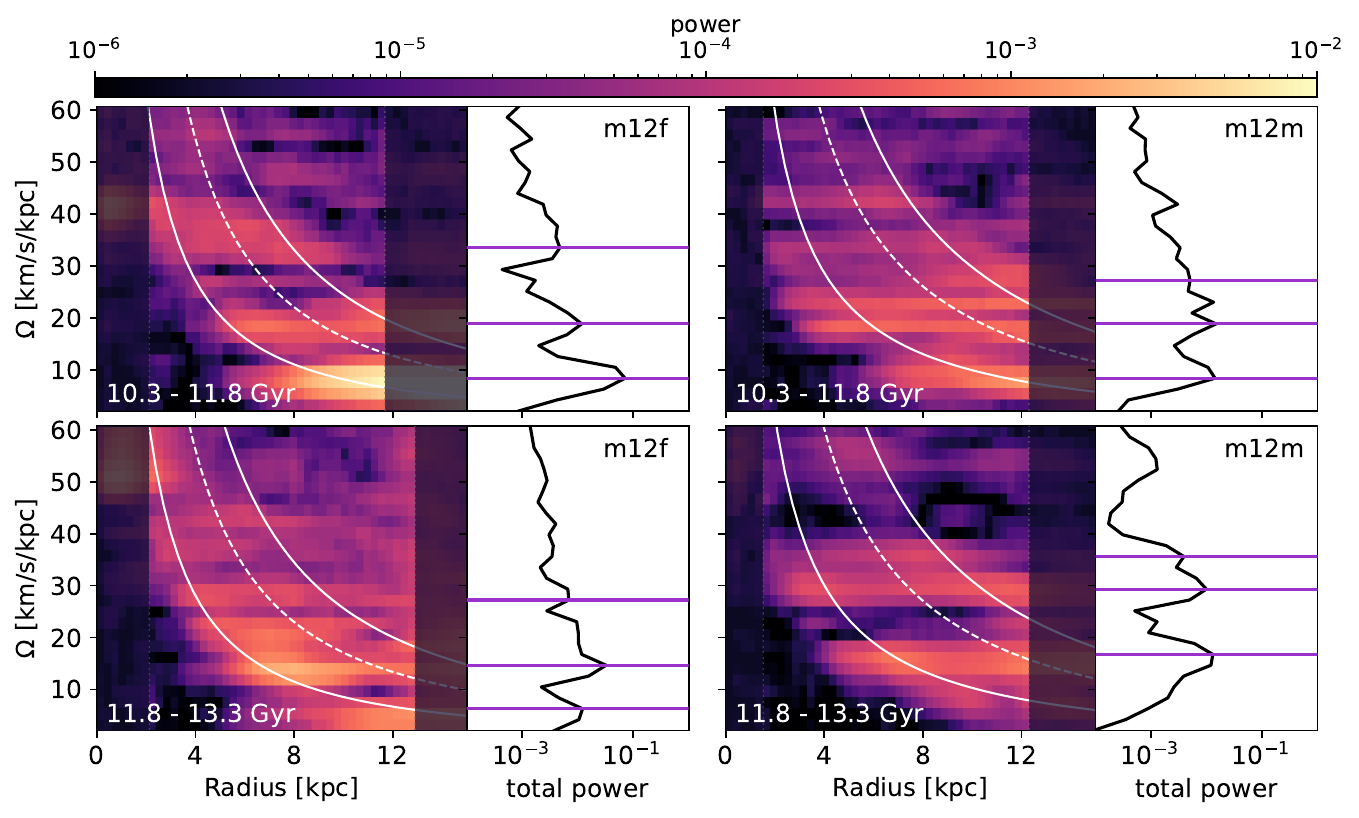}
\caption{Power spectra for m12f (left column) and m12m (right column) at two different time intervals. 
The left side of each plot shows the spectrogram of pattern speeds as a function of radius (Eq.~\ref{eq:6}), colored by power, and the right side shows the radially integrated power (Eq.~\ref{eq:total_power}).
More power indicates stronger $m=2$ spiral structure present over the time baseline.
The darkened rectangles cut out the bar and region beyond the disk and are not included in the integrated spectra.
The Inner Linblad, Outer Linblad, and corotation resonances are plotted as the two solid white lines and dashed white line, respectively.
The purple horizontal lines on the right panels show peaks in the power spectrum and identify the three most dominant pattern speeds.
The structure of $m=2$ spiral amplitudes differs between m12f and m12m; m12f experiences strong tidal interactions whereas m12m evolves secularly in the time baselines shown.
\label{fig:spectra}}
\end{center}
\end{figure*}

We note that this technique is effective for analyzing the dominant spiral modes most important to the dynamics and evolution of the disk.
Thus, more flocculent or locally isolated features, which play a role in the evolution of the galaxy, are not characterized in this paper.
We are now able to construct a power spectrum showing the power and spatial distribution of spiral amplitudes over a given time baseline, shown in Fig. \ref{fig:spectra}. 

Fig.~\ref{fig:spectra} shows the results of applying the WDFT analysis on the $m=2$ multiplicity over two different $1.5$~Gyr time baselines for galaxies m12f and m12m.
The left-hand side of each of the four sets of plots shows the spectrogram of pattern speeds as a function of radii, revealing the spatial distribution of $m=2$ spiral structure. 
The time baseline is in the bottom left corner.
The colors correspond to power, which can be interpreted as proportionally how over-dense the waves propagating at a certain frequency are compared to the background density of stars. 
Higher values of power correspond to a higher over-density of stars.
The dashed white line shows the circular frequency $\Omega_c$ and the solid white lines show $\Omega_c \pm \kappa/2$, respectively. 
These three lines correspond to where the Inner Linblad, and Outer Linblad, and corotation resonances occur for stars on circular or near circular orbits. 
The ILR and OLR are defined as $m(\Omega_{\phi} - \Omega_p) = \pm \kappa$ where $m$ is the pattern multiplicity, $\Omega_p$ is the pattern speed of the density wave, $\Omega_{\phi}$ is the azimuthal frequency of the orbit, and $\kappa$ is the radial frequency of the orbit.
We obtain these curves at the midpoint of each time baseline.
The darkened regions in each spectrogram cover the inner regions where the bar extends to \citep{ansar2025} and beyond the radius enclosing 90\% of the stellar mass, $R_{90}$.
$R_{90}$ is a measure that effectively captures the disk material \citep{bellardini2022} and we interpret this value as the edge of the disk in this paper.
The power in these regions are not considered when calculating the total power.

The right-hand side of each of the four sets of plots in Fig.~\ref{fig:spectra} shows the power spectrum that results from radially integrating the spectrogram.
Peaks in the power spectrum reveal what pattern speeds are most dominant across the disk over a time baseline, and we identify the three most dominant frequencies, indicated by the purple horizontal lines, using scipy.find\_peaks.
Noise contribution to the power spectrum arising from the number of particles in the simulation is minimal and does not affect the identification of the highest amplitude peaks.
We impose a minimum of two times the frequency resolution, $2\Delta\Omega$, between neighboring peaks.

Thus, Fig.~\ref{fig:spectra} shows us, over a $1.5$~Gyr time baseline, where $m=2$ spiral structure is spatially present, how powerful the spiral structure is, and their pattern speeds, $\Omega_p$. 
In each spectrogram, we see the simultaneous presence of multiple spiral arms, displaying the complex dynamics of the disk.
Here we see, for both m12f and m12m, that the properties of spiral structure evolve over the two windows of time shown.

In the top left panel of Fig.~\ref{fig:spectra}, we can connect the higher concentration of power at the outskirts of the disk and lower pattern speed ($\sim10$~km/s/kpc) to the pericentric passage with a satellite galaxy (8:1 mass ratio) that m12f undergoes during this time baseline.
This is the same satellite galaxy passage that is shown in Fig.~\ref{fig:contours}.
A later time baseline is shown in the bottom left panel, during which the satellite galaxy merges with m12f. 
There is more noise present in the spectrogram as the merger event disrupts the disk.

The same two time baselines are shown in the right panels of Fig.~\ref{fig:spectra}, but for m12m.
This galaxy experiences predominantly secular evolution over the time period shown.
From $10.3-11.8$~Gyr we can pick out two dominant frequencies, one at $\sim10$~km/s/kpc and another at $\sim20$~km/s/kpc.
The pattern speed of the structures evolves over time, as we identify the dominant frequencies at $\sim15$~km/s/kpc and $\sim30$~km/s/kpc in bottom right panel.
The evolution of dominant frequencies is discussed in \S~\ref{sec:results}.

While the radial extent of spiral structure is quite extended in both galaxies, the distribution of power radially in m12m is more spread out, and it is difficult to visually identify where the power is most radially concentrated.
In m12f, it is more visually clear where there is more power, and we see that regions of higher power in dominant structures are near the ILR.

See Appendix~\ref{app:spectrogram} for spectrograms of all 10 FIRE-2 galaxies analyzed for one time baseline.
In addition to performing this analysis on the stellar population, we apply the WDFT analysis on stellar populations of varying ages and star forming gas, which we discuss further in \S\ref{ssec:star_forming_gas}. 
To our knowledge, the application of this spectral analysis technique to gas and different populations of stars has not been done before and the results give important insight into how the dominant frequencies vary between different populations.

\section{Results} \label{sec:results}
\begin{figure*}
\begin{center}
\includegraphics[width=0.9\linewidth]{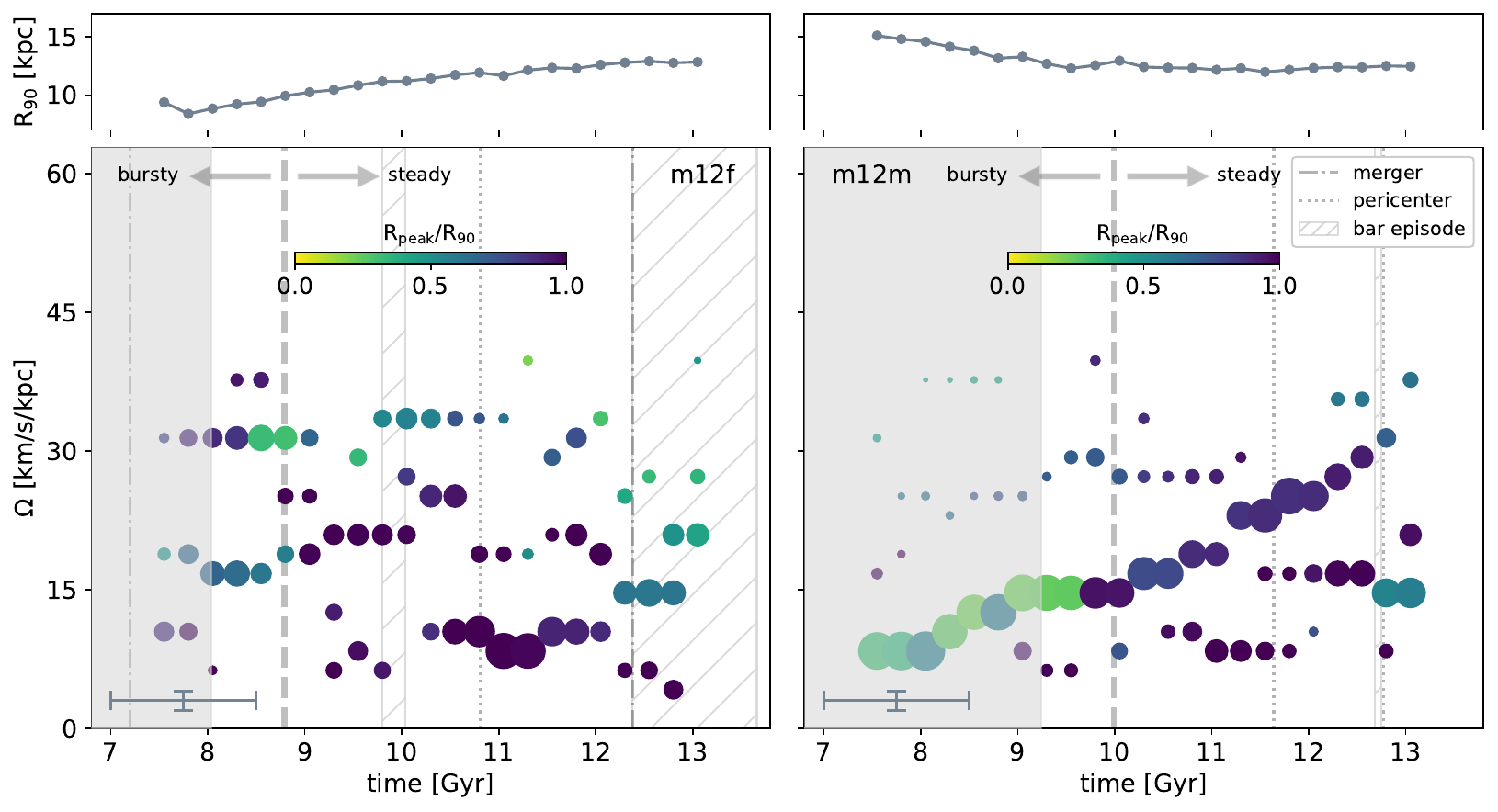}
\caption{Frequency evolution of the three most dominant $m=2$ amplitudes for m12f (left) and m12m (right). 
The top row shows the radius enclosing 90\% of the stellar mass, $R_{90}$, as a function of time. 
Points are color-coded by the radius of peak spiral power relative to $R_{90}$, and the size is proportional to their amplitudes during the relevant time baseline. 
The time baseline (1.5~Gyr) and frequency resolution ($\Delta\Omega_p\sim 2.1$~km/s/kpc) are shown in the bottom left.
The thick gray vertical dashed line indicates when the galaxy transitions from bursty to steady star formation rate.
The grayed out region shows the time period that we do not analyze the galaxy.
We indicate when the galaxy undergoes merger events (dash-dot lines), pericentric passages (dotted line), and bar episodes (hatched rectangle).
The evolution of $m=2$ amplitudes is more segmented in an environment with significant external influence in m12f, compared to a galaxy with primarily secular evolution as m12m. 
Whereas m12m has a single evolving spiral amplitude that is primarily dominant across time, m12f hosts multiple instances of dominant spiral amplitudes scattered at different pattern speeds across time.
This is discussed further in \S\ref{sec:results} and can be seen more clearly in Fig.~\ref{fig:all_sims_identified_spirals}.
\label{fig:bubble_m12f_m12m}}
\end{center}
\end{figure*}

By applying each $1.5$~Gyr WDFT analysis at equally spaced intervals of $0.25$~Gyr from $\sim6.8$~Gyr to present day, we are able to construct the time-evolution of dominant $m=2$ spiral frequencies, shown in the bottom row of Fig.~\ref{fig:bubble_m12f_m12m}. 
The top panel of Fig.~\ref{fig:bubble_m12f_m12m} shows the evolution of $R_{90}$, the radius enclosing 90\% of the stellar mass within the central 20~kpc, as described in \cite{bellardini2022}, calculated at the midpoint of each time baseline.
Each of the three data points for a given time baseline corresponds to one of the three peaks identified in the power spectrum (Fig.~\ref{fig:spectra}).
We remind the reader that we perform this analysis on all stars and begin characterizing the spiral structure at the transition from bursty to steady star formation plus half the time baseline and that times prior to this are blocked out in gray.
We choose to begin our characterization at this time for consistency between the galaxies analyzed.
Points are sized proportional to their amplitudes at a given time baseline and are color coded by radius of peak power, $R_{\text{peak}}$, normalized by $R_{90}$.
$R_{\text{peak}}$ is identified using the spectrograms in Fig.~\ref{fig:spectra}.
Thus, $R_{\text{peak}}/R_{90}$ tells us where in the disk proportionally the peak power occurs and lets us track how the the location of $R_{\text{peak}}$ changes over time.
We note that because the stellar density decreases exponentially with increasing radius, the WDFT analysis will be more sensitive to perturbations at larger radii.
See Appendix~\ref{app:mass_weight} for an alternative approach that normalizes the power spectrum to account for the increased power at larger radii.
This results in more power identified in spiral modes present at larger radii, which systematically pushes the location of $R_{\text{peak}}$ to larger values.
However, despite $R_{\text{peak}}$ being identified at larger radii, it is still a helpful measure to probe which general regions of the disk undergo more spiral activity.

An interesting feature we see present in both galaxies, but more strikingly in m12m, is that $R_{\text{peak}}$ is concentrated in the inner regions of the disk during earlier periods of bursty star formation but quickly moves outward to larger radii as the galaxy transitions to steady star formation rate. 
This trend is present in m12f, but not as pronounced.
We see power move outwards at the transition to steady star formation across many galaxies in our analysis (see Appendix~\ref{app:bubble}).
This outward movement of power may be related to the inside-out formation of disks.
\cite{graf2024} finds that the FIRE-2 galaxies do experience meaningful inside-out growth, particularly after the time that the disks settle.
They find that gas and young stars are particularly affected by this radial inside-out growth, which may be related to  power moving to larger radii around the bursty to steady transition. 
Though we focus on analyzing the galaxies once they are in steady star formation and have not further explored this feature of power propagating outwards in the disk during the bursty to steady transition, it is an interesting characteristic to note.

\begin{figure*}
\begin{center}
\includegraphics[width=\linewidth]{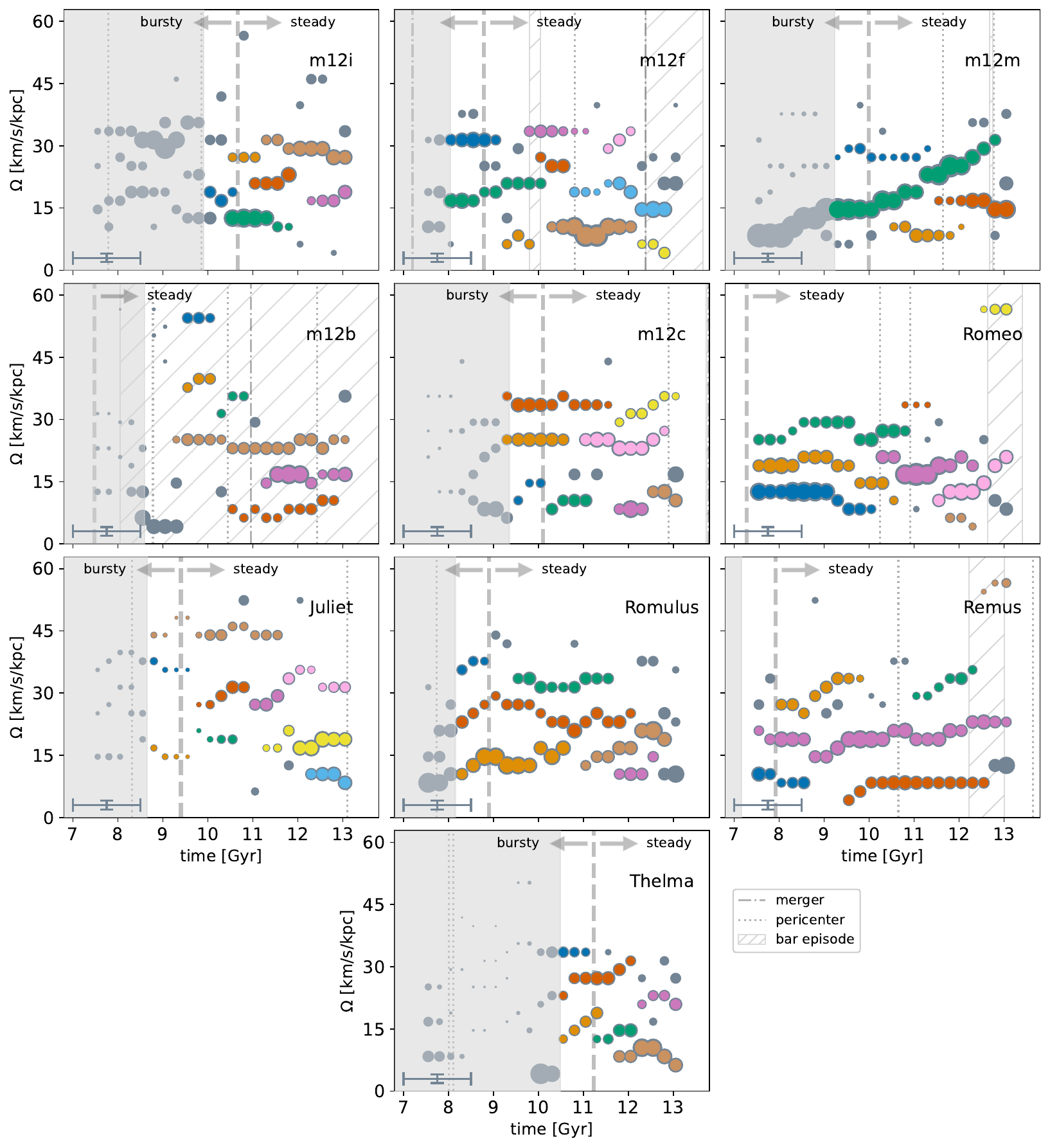}
\caption{Similar to Fig.~\ref{fig:bubble_m12f_m12m}, showing identified episodes of $m=2$ spiral patterns using the procedure described in Appendix~\ref{app:identify_arms} for all FIRE-2 galaxies analyzed.
Each color represents a distinct spiral episode, where a corresponding entry in Table~\ref{tab:spiral-properties} describes its properties. 
Point sizes are proportional to their amplitudes during the relevant time baseline.
Gray points represent $m=2$ spiral amplitudes present over the time baseline, but not identified as a member of a spiral episode.
We see significant diversity in the properties of identified spirals across all galaxies.
\label{fig:all_sims_identified_spirals}}
\end{center}
\end{figure*}

Stellar bars can excite spirals, and while it is difficult to disentangle the specific mechanism driving the dominant amplitudes, especially in cosmological simulations, we note instances where spiral episodes seem to be associated with bar episodes.
In m12f, we see in the bottom left panel of Fig.~\ref{fig:bubble_m12f_m12m} that the bar episode at around $10$~Gyr corresponds to the emergence of a strong, fast moving ($33$~km/s/kpc) structure.
This bar episode is short lived ($0.18$~Gyr) and formed through internal evolution of the galaxy \citep{ansar2025}.
This spiral amplitude fades away after the bar dissipates, suggesting that it may have been excited by the bar.
The bar in m12m is very short lived, but we see in the bottom right panel of Fig.~\ref{fig:bubble_m12f_m12m} the potential impact of the bar as we identify a spiral amplitude with $\Omega_p\sim35$~km/s/kpc emerge around the time of the bar episode.

Interactions with satellite galaxies can also excite spirals, and we see the impact of pericentric passages and merger events on the evolution of dominant spiral amplitudes in our simulations.
In the bottom left panel of Fig.~\ref{fig:bubble_m12f_m12m}, we see that the pericentric passage in m12f around $10.8$~Gyr leads to a slight decrease in $\Omega_p$ and increase in relative power of the most dominant spiral amplitude ($10$~km/s/kpc).
This dominant spiral amplitude persists for multiple Gyr until the galaxy merges with the satellite at $12.4$~Gyr.

In the bottom right panel of Fig.~\ref{fig:bubble_m12f_m12m}, we see that m12m has a single spiral amplitude with relatively more power than other dominant amplitudes and is maintained throughout nearly the entire time of analysis with steadily increasing $\Omega_p$.
As this dominant spiral amplitude increases $\Omega_p$, there is an emergence of another relatively strong spiral frequency that occupies the slower range of $\Omega_p$. 
During the period of analysis, m12m has a quiescent, secularly driven evolution.
And while there are two pericentric passages of satellites at late times, the evolution of m12m's disk does not appear to be strongly impacted by its tidal interactions, as the interacting satellite galaxies are not massive enough to significantly perturb the disk \citep[see discussion in][for further details]{ansar2025}.

See Appendix~\ref{app:bubble} for the time evolution of $m=2$ spiral amplitudes for all 10 FIRE-2 galaxies analyzed in this paper.
Across all galaxies, we see that pattern speeds of dominant spiral modes do not change drastically on short time scales in the absence of satellite interactions.
We find instances where a change in $\Omega_p$ is correlated with a satellite interaction, but this is not always the case as the disk's response to tidal perturbations is dependent on various properties, such as mass and direction of orbit of the satellite galaxy.
We defer an exploration of how changing pattern speeds affects the evolution of the disk to future work.
As expected, pericentric passages can generate high power spiral signals in the outskirts of the disk.
In contrast, bar episodes may move power to more inner regions of the disk and more commonly lead to the presence of dominant spiral modes with faster $\Omega_p$.

Fig.~\ref{fig:all_sims_identified_spirals} shows the resulting $m=2$ spiral episodes identified by connecting the dominant $m=2$ spiral amplitudes coherently across time in all 10 FIRE-2 galaxies.
See Appendix~\ref{app:identify_arms} for a technical description of the algorithm.
Each color represents a separate $m=2$ spiral episode.
Amplitudes prior to when we begin our analysis, as well as those that do not meet the criteria described in Appendix~\ref{app:identify_arms}, are shown as the gray data points.

We focus on the results in Fig.~\ref{fig:all_sims_identified_spirals} for galaxies m12f (top middle panel) and m12m (top right panel) in this section.
There are more spiral episodes identified in m12f than m12m ($9$ vs $4$), and the overall evolution of $m=2$ spirals in m12f appears more segmented than in m12m. 
There is also not a singular episode that dominates the evolution in m12f.
While the pericentric passage in m12f around $10.8$~Gyr provokes an amplitude that is dominant for nearly 1~Gyr, the spiral episodes during all other times are relatively similar in amplitude, indicated by the relative size of each point.
This is in stark contrast to m12m, where there is a single spiral episode (green) that dominates over the other episodes for a majority of its existence. 

Across all galaxies, at any given time, we see multiple prolonged $m=2$ spiral episodes that propagate through the disk at different pattern speeds and that follow different phenomenology.
Most spiral arms stay at relatively fixed pattern speeds over time. 
Generally, departures from this are associated with interaction with larger scale structure (e.g., pericentric passage and bar episodes).

We have described the evolution of the dominant $m=2$ spiral episodes in m12f and m12m, and the subsections below focus on reporting and analyzing the properties and characteristics of spiral episodes across all galaxies in our sample.
The properties of identified $m=2$ spiral episodes for all galaxies are listed in Table~\ref{tab:spiral-properties}.

\subsection{Instances of $m=2$ Spiral Episodes} \label{ssec:instances}

The number of $m=2$ spiral episodes we identify (column~2 of Table~\ref{tab:spiral-stats}) is impacted by the length of time each galaxy is analyzed for.
This length of time is directly determined by when the galaxy transitions from bursty to steady star formation, $t_{\text{B-S}}$, plus $1/2$ the length of the time baseline used ($(1/2)S\Delta t = 0.75$~Gyr). 
Because each galaxy has a different transition time, rather than directly comparing the number of spiral episodes in each galaxy, we measure the rate of $m=2$ spiral episodes, which we sometimes simply call the ``rate" (column~3 of Table~\ref{tab:spiral-stats}).
This is the number of $m=2$ spiral episodes divided by the length of time under consideration.
Using this definition of rate, we find a median of 1.35 $m=2$ spiral episodes per Gyr of analysis across all galaxies.
That is, there is roughly 1.35 $m=2$ spiral episodes that occur per Gyr. 
Multiple $m=2$ spirals exist simultaneously at any given time, with a median lifetime of 1.56~Gyr across all galaxies.
However, the individual lifetimes of these arms vary markedly and is discussed in \S\ref{ssec:duration}.
We note that Table~\ref{tab:spiral-stats} contains both the mean and median values of each column; however, for brevity, we report only the median in the text.

Thelma, m12c, and Juliet host the highest rates, at 1.81~Gyr$^{-1}$, 1.80~Gyr$^{-1}$, and 1.75~Gyr$^{-1}$, respectively.
Thelma and m12c transition to steady star formation more recently than most other galaxies analyzed, 2.7~Gyr and 3.7~Gyr ago, and it is interesting that we find higher rates in more recently settled disks.
The higher rates in these galaxies are in part because many of their spiral episodes are short-lived, which can be seen in Fig.~\ref{fig:all_sims_identified_spirals}.
We observe that many instances of spiral episodes recurring at similar pattern speeds across time, but identified as distinct from each other.
An example of this can be found in m12c at $\sim25$~km/s/kpc, where the orange and pink episodes are separated by a single time baseline that this spiral frequency is not dominant in the disk.
We note that instances like this may affect the resulting rate of spiral episodes.
Thelma and Juliet do not undergo any tidal interactions over the time of analysis and are both kinematically hot among the galaxies we analyze.
Juliet is also the lowest mass galaxy from our sample.
m12c experiences tidal influence from a satellite interaction at 12.89~Gyr on a relatively prograde orbit and hosts a bar near present day \citep{ansar2025}.
However, this tidal interaction and bar episode occur at the end of its evolution, and, like Thelma and Juliet, m12c is kinematically hot during most of its lifetime.

m12m and Remus have the lowest rates at 0.88~Gyr$^{-1}$ and 0.90~Gyr$^{-1}$, respectively.
These galaxies host longer-lived spiral episodes and have kinematically colder disks.
While m12m is the most massive galaxy in our sample, Remus is among the lowest mass galaxy.
Neither of these galaxies experience strong tidal interactions over the time of analysis, although both galaxies host a bar towards the end of their evolution.

The LG-like pairs generally experience an earlier transition to steady star formation than the isolated galaxies and have a slightly higher median rate in comparison (1.10~Gyr$^{-1}$ vs 1.54~Gyr$^{-1}$).

\subsection{Spiral Pattern Speeds ($\Omega_p$)} \label{ssec:pattern_speeds}

We see a large variety in the range of $\Omega_p$ of $m=2$ spiral episodes, both within each galaxy and across all galaxies.
The minimum, maximum, and median $\Omega_p$ across galaxies is shown in columns 4-6 of Table.~\ref{tab:spiral-stats}.
We find the median $\Omega_{p,\text{min}}=7.3$ and $\Omega_{p,\text{max}}=36.7$~km/s/kpc across all galaxies.
All galaxies in our sample have a similar range of minimum and maximum $\Omega_p$.
$\Omega_{p,\text{min}}$ ranges between $5-10$~km/s/kpc, and $\Omega_{p,\text{max}}$ has slightly more variation, ranging between 30-55~km/s/kpc.
In general, the range of $\Omega_p$ is comparable between the LG-like pairs and isolated galaxies, and the median $\Omega_p$ across all galaxies is $\Omega_{p,\text{median}}=19.6$~km/s/kpc.

\clearpage
\pagebreak

\startlongtable
\begin{deluxetable*}{cr>{\hspace{-7pt}- }lr>{\hspace{-7pt}- }lr>{\hspace{-7pt}- }lccccccccc}
\vspace*{-1cm}
\tablecaption{Properties of identified $m=2$ spiral episodes.\label{tab:spiral-properties}}
\tablehead{\vspace{-0.1cm}
  Sim & \multicolumn{2}{c}{Time} & \multicolumn{2}{c}{Radii} & \multicolumn{2}{c}{$\Omega_p$} & \colhead{$a_{2\text{,max}}$} & \colhead{$\text{time}_{\text{max}}$} & \colhead{$R_{\text{peak}}$} & \colhead{$R_{\text{ILR}}$} & \colhead{$R_{\text{CR}}$} & \colhead{$R_{\text{OLR}}$} & \colhead{$\Delta\text{time}$} & \colhead{$\Delta R$} & \colhead{Nearest} \vspace{-0.2cm} \\
   & & \colhead{} & \colhead{} & \colhead{} & \colhead{} & \colhead{} & \colhead{} & & & & & & & & \colhead{Resonance} \vspace{-0.4cm} \\
  & \multicolumn{2}{c}{[Gyr]} & \multicolumn{2}{c}{[kpc]} & \multicolumn{2}{c}{[km s$^{-1}$kpc$^{-1}$]}\hspace{-5pt} & & [Gyr] & [kpc] & [kpc] & [kpc] & [kpc] & [Gyr] & [kpc] & 
}
\startdata
\multirow{6}{*}{m12i} & 9.80 & 10.80 & 3.0 & 7.2 & 16.8 & 18.8 & 0.061 & 10.19 & 4.8 & 5.1 & 8.7 & 11.4 & 1.00 & 4.2 & ILR \\
  & 10.30 & 11.30 & 2.7 & 8.4 & 27.2 & 27.2 & 0.030 & 10.89 & 8.1 & 3.9 & 6.6 & 9.0 & 1.00 & 5.7 & OLR \\
  & 10.30 & 12.05 & 4.2 & 8.4 & 10.5 & 12.6 & 0.069 & 10.87 & 8.1 & 6.9 & 11.4 & - & 1.75 & 4.2 & ILR \\
  & 10.80 & 12.05 & 3.3 & 9.0 & 20.9 & 23.0 & 0.054 & 11.65 & 9.0 & 4.8 & 8.1 & 10.8 & 1.25 & 5.7 & CR \\
  & 12.05 & 13.30 & 5.4 & 9.3 & 16.8 & 18.8 & 0.045 & 12.87 & 9.3 & 6.0 & 9.9 & 13.2 & 1.25 & 3.9 & CR \\
  & 11.05 & 13.30 & 2.1 & 8.6 & 27.2 & 31.4 & 0.043 & 11.19 & 3.9 & 3.6 & 6.3 & 8.4 & 2.25 & 6.5 & ILR \\ \hline
\multirow{9}{*}{m12f} & 7.80 & 9.30 & 2.1 & 9.4 & 31.4 & 31.4 & 0.050 & 8.58 & 3.0 & 3.3 & 5.7 & 7.8 & 1.50 & 7.3 & ILR \\
  & 9.05 & 10.05 & 6.3 & 10.8 & 6.3 & 8.4 & 0.070 & 9.58 & 10.5 & 11.7 & - & - & 1.00 & 4.5 & ILR \\
  & 7.80 & 10.30 & 3.3 & 9.0 & 16.8 & 20.9 & 0.083 & 8.16 & 5.7 & 4.8 & 8.1 & 10.5 & 2.50 & 5.7 & ILR \\
  & 9.80 & 10.80 & 2.7 & 11.7 & 25.1 & 27.2 & 0.058 & 10.49 & 11.1 & 4.2 & 7.2 & 9.6 & 1.00 & 9.0 & OLR \\
  & 9.55 & 11.30 & 2.1 & 11.2 & 33.5 & 33.5 & 0.033 & 10.04 & 6.3 & 3.3 & 5.7 & 7.8 & 1.75 & 9.1 & CR \\
  & 10.05 & 12.30 & 5.4 & 12.0 & 8.4 & 10.5 & 0.192 & 11.11 & 12.0 & 8.4 & 13.8 & - & 2.25 & 6.6 & CR \\
  & 11.30 & 12.30 & 2.4 & 12.3 & 29.3 & 33.5 & 0.041 & 11.84 & 9.3 & 3.6 & 6.3 & 8.4 & 1.00 & 9.9 & OLR \\
  & 12.05 & 13.05 & 6.0 & 12.9 & 4.2 & 6.3 & 0.079 & 11.55 & 12.9 & 12.3 & - & - & 1.00 & 6.9 & ILR \\
  & 10.55 & 13.05 & 4.8 & 12.9 & 14.7 & 20.9 & 0.098 & 12.45 & 7.8 & 5.4 & 9.3 & 12.6 & 2.50 & 8.1 & CR \\ \hline
\multirow{4}{*}{m12m} & 9.05 & 11.55 & 2.1 & 12.3 & 27.2 & 29.3 & 0.049 & 9.56 & 9.0 & 3.9 & 7.2 & 10.2 & 2.50 & 10.2 & OLR \\
  & 10.30 & 12.30 & 6.6 & 12.0 & 8.4 & 10.5 & 0.078 & 10.72 & 12.0 & 11.4 & - & - & 2.00 & 5.4 & ILR \\
  & 9.05 & 13.05 & 1.5 & 12.3 & 14.7 & 31.4 & 0.170 & 9.19 & 3.0 & 5.4 & 9.6 & 12.9 & 4.00 & 10.8 & ILR \\
  & 11.30 & 13.30 & 4.2 & 12.3 & 14.7 & 16.8 & 0.064 & 12.38 & 12.3 & 6.6 & 11.7 & - & 2.00 & 8.1 & CR \\ \hline
\multirow{6}{*}{m12b} & 9.30 & 10.30 & 2.1 & 6.7 & 54.5 & 54.5 & 0.041 & 9.76 & 2.4 & 2.4 & 4.2 & 5.7 & 1.00 & 4.6 & ILR \\
  & 9.30 & 10.30 & 2.7 & 6.6 & 37.7 & 39.8 & 0.048 & 8.94 & 6.6 & 3.3 & 5.4 & 6.9 & 1.00 & 3.9 & OLR \\
  & 10.05 & 11.05 & 2.7 & 8.1 & 31.4 & 35.6 & 0.078 & 11.01 & 7.5 & 3.6 & 6.0 & 8.1 & 1.00 & 5.4 & OLR \\
  & 10.30 & 13.05 & 8.4 & 10.5 & 6.3 & 10.5 & 0.168 & 10.82 & 10.5 & 10.2 & - & - & 2.75 & 2.1 & ILR \\
  & 11.05 & 13.30 & 2.1 & 10.0 & 14.7 & 16.8 & 0.104 & 11.50 & 8.1 & 6.3 & 10.5 & 14.1 & 2.25 & 7.9 & ILR \\
  & 9.05 & 13.30 & 2.1 & 10.8 & 23.0 & 25.1 & 0.116 & 10.57 & 9.9 & 4.5 & 7.5 & 10.2 & 4.25 & 8.7 & OLR \\ \hline
\multirow{8}{*}{m12c} & 9.30 & 10.30 & 3.6 & 9.1 & 10.5 & 14.7 & 0.058 & 10.59 & 8.1 & 5.7 & 9.6 & 12.9 & 1.00 & 5.5 & CR \\
  & 9.05 & 10.80 & 1.5 & 9.0 & 25.1 & 25.1 & 0.046 & 9.91 & 7.8 & 3.6 & 6.3 & 8.7 & 1.75 & 7.5 & OLR \\
  & 10.05 & 11.30 & 5.7 & 9.0 & 8.4 & 10.5 & 0.056 & 10.79 & 9.0 & 7.5 & 12.3 & - & 1.25 & 3.3 & ILR \\
  & 9.05 & 11.80 & 1.5 & 8.4 & 33.5 & 35.6 & 0.076 & 10.04 & 1.5 & 2.7 & 5.1 & 6.9 & 2.75 & 6.9 & ILR \\
  & 11.55 & 12.55 & 5.4 & 9.6 & 8.4 & 8.4 & 0.096 & 12.15 & 9.6 & 9.3 & - & - & 1.00 & 4.2 & ILR \\
  & 12.30 & 13.30 & 4.8 & 9.9 & 10.5 & 12.6 & 0.082 & 12.36 & 9.0 & 6.9 & 11.4 & - & 1.00 & 5.1 & ILR \\
  & 10.80 & 13.05 & 2.1 & 9.9 & 23.0 & 27.2 & 0.083 & 11.55 & 9.6 & 3.9 & 6.6 & 9.0 & 2.25 & 7.8 & OLR \\
  & 11.55 & 13.30 & 1.5 & 9.1 & 29.3 & 35.6 & 0.039 & 11.36 & 7.5 & 3.0 & 5.4 & 7.5 & 1.75 & 7.6 & OLR \\
\tablebreak
\multirow{8}{*}{Romeo} & 7.30 & 10.30 & 8.1 & 11.7 & 8.4 & 12.6 & 0.113 & 8.83 & 11.7 & 6.6 & 11.4 & - & 3.00 & 3.6 & CR \\
  & 7.30 & 10.80 & 3.3 & 11.1 & 10.5 & 20.9 & 0.112 & 8.73 & 11.1 & 4.8 & 8.4 & 11.4 & 3.50 & 7.8 & OLR \\
  & 7.30 & 11.05 & 2.4 & 10.7 & 25.1 & 29.3 & 0.049 & 8.00 & 9.6 & 3.6 & 6.0 & 8.4 & 3.75 & 8.3 & OLR \\
  & 10.55 & 11.55 & 7.2 & 7.8 & 33.5 & 33.5 & 0.018 & 10.25 & 7.8 & 3.0 & 5.4 & 7.5 & 1.00 & 0.6 & OLR \\
  & 10.05 & 12.55 & 4.2 & 12.8 & 16.8 & 20.9 & 0.054 & 10.68 & 12.6 & 5.1 & 8.7 & 11.7 & 2.50 & 8.6 & OLR \\
  & 11.55 & 12.55 & 9.6 & 12.6 & 4.2 & 6.3 & 0.039 & 12.06 & 12.6 & 12.0 & - & - & 1.00 & 3.0 & ILR \\
  & 11.30 & 13.30 & 6.3 & 12.6 & 10.5 & 20.9 & 0.058 & 12.35 & 9.0 & 6.9 & 12.0 & - & 2.00 & 6.3 & ILR \\
  & 12.30 & 13.30 & 1.5 & 6.0 & 56.5 & 56.5 & 0.022 & 12.73 & 1.5 & 1.8 & 3.6 & 5.1 & 1.00 & 4.5 & ILR \\ \hline
\multirow{9}{*}{Juliet} & 8.55 & 9.80 & 1.5 & 4.2 & 35.6 & 37.7 & 0.040 & 8.91 & 1.8 & 3.0 & 5.1 & 6.9 & 1.25 & 2.7 & ILR \\
  & 8.55 & 9.80 & 3.9 & 6.6 & 14.7 & 16.8 & 0.042 & 9.09 & 5.4 & 6.0 & 10.2 & 13.8 & 1.25 & 2.7 & ILR \\
  & 9.55 & 10.80 & 3.6 & 3.9 & 18.8 & 20.9 & 0.023 & 9.04 & 3.6 & 4.8 & 8.4 & 11.4 & 1.25 & 0.3 & ILR \\
  & 9.55 & 11.05 & 2.7 & 7.2 & 27.2 & 31.4 & 0.025 & 11.52 & 6.3 & 3.6 & 6.3 & 8.7 & 1.50 & 4.5 & CR \\
  & 10.80 & 11.80 & 3.3 & 6.9 & 27.2 & 29.3 & 0.019 & 11.50 & 6.3 & 3.9 & 6.6 & 9.0 & 1.00 & 3.6 & CR \\
  & 8.55 & 11.80 & 1.8 & 5.4 & 44.0 & 48.2 & 0.028 & 8.81 & 2.7 & 2.4 & 4.2 & 6.0 & 3.25 & 3.6 & ILR \\
  & 11.55 & 13.30 & 5.7 & 6.3 & 31.4 & 35.6 & 0.023 & 11.60 & 6.3 & 3.3 & 5.7 & 8.1 & 1.75 & 0.6 & CR \\
  & 11.05 & 13.30 & 4.8 & 8.7 & 16.8 & 20.9 & 0.058 & 12.23 & 8.7 & 5.4 & 9.3 & 12.6 & 2.25 & 3.9 & CR \\
  & 12.05 & 13.30 & 5.7 & 9.0 & 8.4 & 10.5 & 0.034 & 12.13 & 8.4 & 8.1 & 13.8 & - & 1.25 & 3.3 & ILR \\ \hline
\multirow{6}{*}{Romulus} & 8.05 & 9.05 & 2.4 & 9.3 & 35.6 & 37.7 & 0.045 & 8.53 & 3.0 & 3.0 & 5.4 & 7.5 & 1.00 & 6.9 & ILR \\
  & 8.05 & 10.80 & 0.3 & 10.8 & 10.5 & 16.8 & 0.077 & 8.48 & 8.1 & 6.9 & 11.7 & - & 2.75 & 10.5 & ILR \\
  & 9.30 & 11.80 & 3.6 & 11.7 & 31.4 & 33.5 & 0.033 & 10.87 & 9.6 & 3.6 & 6.0 & 8.4 & 2.50 & 8.1 & OLR \\
  & 8.05 & 12.30 & 3.3 & 11.3 & 20.9 & 29.3 & 0.091 & 8.37 & 11.1 & 4.2 & 7.2 & 9.9 & 4.25 & 8.0 & OLR \\
  & 11.55 & 12.80 & 7.5 & 13.5 & 10.5 & 14.7 & 0.067 & 12.92 & 12.6 & 9.0 & - & - & 1.25 & 6.0 & CR \\
  & 10.80 & 13.05 & 4.2 & 14.0 & 12.6 & 20.9 & 0.069 & 12.79 & 13.8 & 6.3 & 10.8 & 14.4 & 2.25 & 9.8 & OLR \\ \hline
\multirow{6}{*}{Remus} & 7.30 & 8.80 & 5.1 & 9.9 & 8.4 & 10.5 & 0.040 & 8.53 & 7.2 & 8.1 & 13.2 & - & 1.50 & 4.8 & ILR \\
  & 7.80 & 10.05 & 2.1 & 10.0 & 25.1 & 33.5 & 0.047 & 8.43 & 9.0 & 3.0 & 5.1 & 6.9 & 2.25 & 7.9 & OLR \\
  & 10.80 & 12.55 & 2.4 & 10.9 & 29.3 & 35.6 & 0.021 & 10.85 & 10.5 & 2.7 & 5.1 & 6.9 & 1.75 & 8.5 & OLR \\
  & 9.30 & 12.80 & 8.7 & 10.8 & 4.2 & 8.4 & 0.055 & 10.06 & 10.8 & 8.1 & 13.5 & - & 3.50 & 2.1 & ILR \\
  & 7.30 & 13.30 & 3.0 & 9.9 & 14.7 & 23.0 & 0.066 & 8.37 & 9.9 & 4.2 & 7.2 & 9.9 & 6.00 & 6.9 & OLR \\
  & 12.30 & 13.30 & 1.8 & 3.3 & 54.5 & 56.5 & 0.022 & 12.99 & 1.8 & 1.8 & 3.3 & 4.5 & 1.00 & 1.5 & ILR \\ \hline
\multirow{6}{*}{Thelma} & 10.30 & 11.30 & 4.8 & 11.1 & 33.5 & 33.5 & 0.022 & 10.60 & 6.9 & 3.0 & 5.4 & 7.5 & 1.00 & 6.3 & OLR \\
  & 10.30 & 11.55 & 4.8 & 11.0 & 12.6 & 18.8 & 0.040 & 11.26 & 6.6 & 5.7 & 9.9 & 13.5 & 1.25 & 6.2 & ILR \\
  & 11.05 & 12.30 & 4.8 & 11.1 & 12.6 & 14.7 & 0.057 & 12.19 & 11.1 & 6.6 & 11.4 & - & 1.25 & 6.3 & CR \\
  & 10.30 & 12.30 & 1.2 & 11.0 & 23.0 & 31.4 & 0.041 & 11.33 & 9.9 & 3.6 & 6.6 & 9.0 & 2.00 & 9.8 & OLR \\
  & 12.05 & 13.30 & 2.4 & 11.5 & 20.9 & 23.0 & 0.028 & 12.56 & 9.9 & 4.5 & 8.1 & 10.8 & 1.25 & 9.1 & OLR \\
  & 11.55 & 13.30 & 7.5 & 11.7 & 6.3 & 10.5 & 0.125 & 12.63 & 11.7 & 9.9 & - & - & 1.75 & 4.2 & ILR
\enddata
\vspace{-0.15cm}
\tablecomments{Columns: (1) Name of the galaxy. 
(2) Time: start and end time of the spiral episode, listed at the midpoint of the start and end time baselines $\mp 0.25$~Gyr. 
(3) Radii: Radial range of the spiral episode. 
(4) $\Omega_p$: Pattern speed range of the spiral episode. 
(5) $a_{2,\text{max}}$: Maximum $a_2$ amplitude of the frequency filtered inverse Fourier transform (see \S\ref{app:identify_arms}). 
(6) Time$_{\text{max}}$: Time of $a_{2\text{,max}}$. 
(7) $R_{\text{peak}}$: Radius at which we identify $a_{2\text{,max}}$. 
(8) $R_{\text{ILR}}$: Radius of Inner Linblad resonance at Time$_{\text{max}}$. 
(9) $R_{\text{CR}}$: Corotation radius at Time$_{\text{max}}$. 
(10) $R_{\text{OLR}}$: Radius of Outer Linblad resonance Time$_{\text{max}}$. 
Blank entries indicate that the resonance radius is beyond the edge of the disk. 
(11) $\Delta\text{time}$: Duration of the spiral episode, calculated from column 2. 
(12) $\Delta R$: Radial extent of the spiral episode. 
(13) Resonance point nearest to $R_{\text{peak}}$.}
\end{deluxetable*}

\clearpage
\pagebreak

\begin{figure*}
\begin{center}
\includegraphics[width=\linewidth]{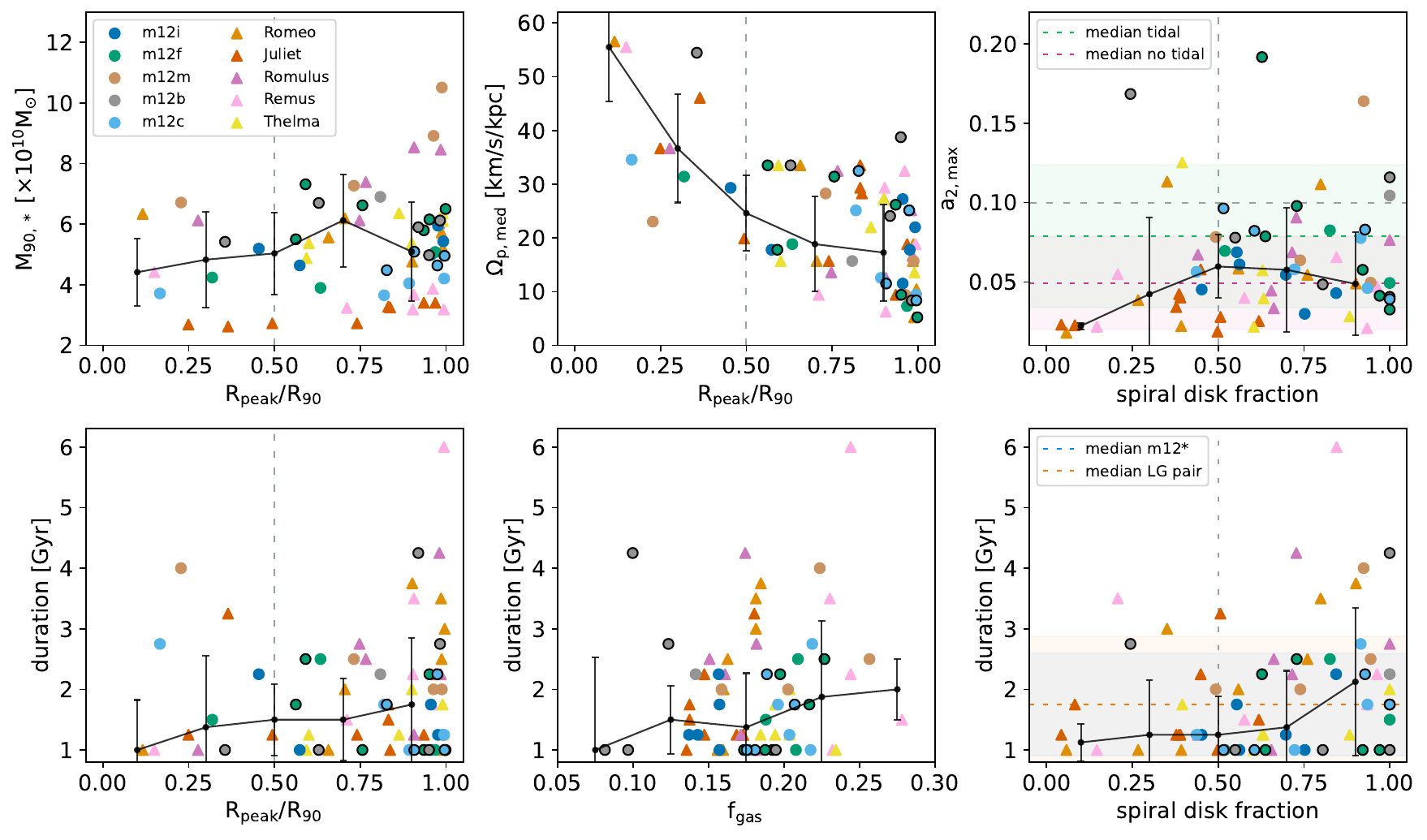}
\caption{Relationships between properties of $m=2$ spiral episodes and galactic environment.
Data points outlined in black indicate a strong tidal interaction (pericentric passage or merger event) occurs during the lifetime of the spiral episode.
Top left: $M_{90,*}$ vs $R_{\text{peak}}/R_{90}$; stellar mass vs radius of spiral episode peak power normalized by $R_{90}$.
$M_{90,*}$ is calculated at the time baseline midpoint nearest $\text{Time}_{\text{max}}$.
Top middle: $\Omega_{p,\text{med}}$ (median pattern speed) vs $R_{\text{peak}}/R_{90}$.
Top right: $a_{2,\text{max}}$ vs spiral disk fraction; maximum amplitude vs fraction of the disk occupied by the spiral episode. 
The horizontal gray dashed line indicates the cutoff for `strong' spiral amplitudes, $a_{2,\text{max}}>0.1$.
The median $a_{2,\text{max}}$ of episodes that do (green) and do not (purple) experience strong tidal interactions are shown by the horizontal dashed lines with the $\pm1 \sigma$ standard deviation shaded.
Bottom left: $\Delta\text{time}$ (duration of the spiral episode) vs $R_{\text{peak}}/R_{90}$.
Bottom middle: $\Delta\text{time}$ vs $f_{\text{gas}}$.
Bottom right: $\Delta\text{time}$ vs spiral disk fraction.
The median $\Delta\text{time}$ for the isolated (blue) and LG-like pairs (orange) are shown by the horizontal dashed lines with the $\pm1 \sigma$ standard deviation shaded.
The top left, top middle, and bottom left panels show a vertical gray dashed line at $R_{\text{peak}}/R_{90}=0.5$; most spiral episodes are above this.
The top right and bottom right panels show a vertical gray dashed line at spiral disk fraction $=0.5$.
In each panel, we divide the data into 5 equally spaced bins and plot the median in each bin as solid black points with errors bars showing the $\pm1 \sigma$ standard deviation.
Tidal interactions play a significant role in generating particularly strong spiral episodes that span across more than half of the disk, but these episodes are not generally very long-lived.
Longer-lived spirals generally occupy large fractions of the disk ($>50$\%), whereas shorter-lived spirals occupy a wider range of disk fractions.
As opposed to the LG-like pairs, the isolated simulations tend to host stronger spirals that occupy larger disk fractions.
Both the LG-like pairs and the isolated galaxies have the same median lifetimes.
\label{fig:spiral_prop_6panel}}
\end{center}
\end{figure*}

Interestingly, Juliet is the lowest mass galaxy we analyze and has a physically small disk at present day (9.6~kpc), but spans a notably large range of $\Omega_p$ and hosts a relatively long-lived (3.25~Gyr), fast ($\Omega_{p,\text{max}}=48.2$~km/s/kpc) spiral episode (Fig.~\ref{fig:all_sims_identified_spirals}, light brown).

m12b, Romeo, and Remus host spiral episodes with the highest $\Omega_{p,\text{max}}$.
The spiral episodes with high $\Omega_p$ in Romeo and Remus may have arisen as a direct result of the bar, as these spirals appear during bar episodes and at high $\Omega_p$ (see Fig.~\ref{fig:all_sims_identified_spirals}).
We also find a clear signature of a bar in the $m=2$ spectrograms for the corresponding time baselines.
\cite{ansar2025} estimates $\Omega_{p,\text{bar}}=69.37$~km/s/kpc in Romeo using the Tremaine-Weinberg method and finds $\Omega_{p,\text{bar}}=93.8$~km/s/kpc in Remus directly calculated using high cadence snapshots.
Since the bar pattern speed in Romeo and Remus are much higher than what we are able to probe given our Nyquist frequency, it is possible that this high $\Omega_p$ amplitude is power extending from the bar. 
Given this caveat, we decide to include these episodes potentially related to the bar in Romeo and Remus in our analysis.
We investigate the spectrograms corresponding to the high $\Omega_p$ episode in m12b and do not find that this amplitude is clearly connected to the signature of a bar, though m12b does host a bar over this time.

We also note an interesting trend of increasing $\Omega_p$ in faster patterns leading up to a bar episode.
Instances of this are found in m12f (light pink), m12m (green), m12c (yellow), Romeo (yellow), and Remus (light brown) (colors refer to spiral episodes in Fig.~\ref{fig:all_sims_identified_spirals}).
The corresponding spectrograms for these galaxies show that more power is present at higher $\Omega_p$ and small radii, which may be driven by the bar or driving the formation of the bar. 
We leave an investigation of the impact of bars on spiral pattern speeds in FIRE-2 simulations to future work.

The spiral episodes with the slowest $\Omega_p$ are in m12f ($\Omega_p=4.2-6.3$~km/s/kpc), Romeo ($\Omega_p=4.2-6.3$~km/s/kpc), and Remus ($\Omega_p=4.2-8.4$~km/s/kpc).
In m12f, this slow spiral may have arisen as a result of a satellite merger.
In Romeo, the slow pattern emerges secularly, is short-lived (1.00~Gyr), and spans the outskirts of the disk (9.6-12.6~kpc).
The slow pattern in Remus is also near the edge of the disk (8.7-10.8~kpc), but is much longer-lived (3.50~Gyr).

The galaxies hosting the fastest $\Omega_p$ are m12b ($\Omega_p=54.5$~km/s/kpc), Romeo ($\Omega_p=56.5$~km/s/kpc), and Remus ($\Omega_p=54.5-56.5$~km/s/kpc).
As discussed earlier in this section, the spirals in Romeo and Remus may be connected to the bar, either as a bar-driven spiral or possibly the bar itself.
All three episodes are located closer to the center of the disk, are short-lived, and have relatively low $a_{2,\text{max}}$.

The median range of $\Omega_p$ spanned by an individual spiral episode (column~4 of Table~\ref{tab:spiral-properties}) is 2.1~km/s/kpc.
This means our spiral patterns largely do not experience significant change in $\Omega_p$ over their lifetimes, and any given spiral occupies only a small range of $\Omega_p$.
We find one significant outlier in this statistic; m12m hosts a spiral with $\Delta\Omega_p=16.7$~km/s/kpc ($\Omega_p=14.7-31.4$~km/s/kpc). 
Notably, this is the most massive galaxy in our sample and evolves generally secularly over the time of analysis, experiencing two pericentric passages that do not strongly influence the disk.
This spiral episode is also the highest amplitude spiral and is long-lived ($4.00$~Gyr), slowly increasing in $\Omega_p$ over its entire lifetime.
Additionally, it occupies a large region of the disk (10.8~kpc) and has a high amplitude ($a_{2,\text{max}}=0.170$) in the absence of strong tidal influence, making it one of the more characteristically grand design spirals we find.

Fig.~\ref{fig:spiral_prop_6panel} shows the relationship between galactic environment and properties of m=2 spiral episodes.
The top middle panel of Fig.~\ref{fig:spiral_prop_6panel} shows median pattern speed, $\Omega_{p,\text{med}}$, versus the radius of peak power of the spiral mode normalized by $R_{90}$, $R_{\text{peak}}/R_{90}$.
Larger $R_{\text{peak}}/R_{90}$ means that a spiral experiences its peak power further out in the disk. 
We note that the WDFT analysis is more sensitive to perturbations at larger radii because of decreasing stellar surface density with radius.
As a result, $R_{\text{peak}}$ will be identified at a larger radius for a given spiral mode radial range.
Even with this bias, there is a clear trend of decreasing $\Omega_p$ with $R_{90}$, $R_{\text{peak}}/R_{90}$.
This matches expectations that pattern speed is radially dependent and is consistent with our $m=2$ spectrograms (Fig.~\ref{fig:spectra}) that show power broadly following the rotation curve.

\subsection{Duration of $m=2$ Spiral Episodes ($\Delta\text{time}$)} \label{ssec:duration}

This section discusses the duration of $m=2$ spiral episodes in our simulations, which we find persist on Gyr timescales.
We emphasize that the lifetime of any individual arm may be much shorter-lived, but the presence of spiral modes with slowly evolving pattern speeds is on Gyr timescales.
We remind the reader that the minimum duration we can identify spiral episodes for is $1.00$~Gyr (see Appendix~\ref{app:identify_arms}).

The overall median duration of spiral episodes is $1.56$~Gyr (Table~\ref{tab:spiral-stats}; column 7).
Remus hosts the longest lived spiral episode ($\Delta\text{time}=6.00$~Gyr).
This galaxy does not have a particularly high stellar mass, but its disk is kinematically cool, which may enable longer-lived spirals.
This particularly long-lived episode emerges we begin our analysis and persists for the entire duration of this galaxy's evolution, as can be seen in Fig.~\ref{fig:all_sims_identified_spirals}.
The pattern speed steadily increases over its lifetime, during which there is a pericentric passage from a satellite galaxy.
This tidal interaction does not strongly impact the disk, but the time of pericenter seems to coincide with a temporary slight increase in $\Omega_p$. 

There are long-lived spiral episodes in m12b (4.25~Gyr), Romulus (4.25~Gyr), and m12m (4.00~Gyr) as well.
Romulus and m12m evolve relatively secularly over the time of analysis, with no strong tidal interactions disrupting the disk.
Additionally, Romulus and m12m are the two most massive galaxies in our sample at $z=0$ and both have relatively kinematically cool disks.
While m12b hosts a relatively high mass disk, it is still surprising that there is such a long-lived spiral episode given the multiple strong tidal interactions m12b experiences during the existence of the long-lived pattern, including a satellite merger.
Despite these perturbations to the disk, the evolution of dominant spiral patterns stays relatively consistent (see Fig.~\ref{fig:all_sims_identified_spirals}).

We find spiral episodes with the minimum identifiable lifetime of $1.00$~Gyr in all galaxies except for m12m.
Given the majority of our galaxies have episodes that persist for the minimum identifiable time, it is reasonable to expect that these galaxies may have strong $m=2$ spiral modes timescales shorter than 1.00~Gyr.
However, most spiral episodes persist for $>$1.00~Gyr, so we are still able to assess the globally dominant $m=2$ spiral features present over the lifetime of the disk.

There is considerable spread in median lifetimes, with a minimum of $\Delta\text{time}_{\text{median}}=1.25$~Gyr in m12i, Juliet, and Thelma and maximum $\Delta\text{time}_{\text{median}}=2.38$~Gyr in Romulus.
The isolated galaxies and LG-like pairs both have $\Delta\text{time}_{\text{median}}=1.75$~Gyr, but the former group has a slightly shorter average duration (1.77~Gyr vs 2.02~Gyr).

The bottom middle panel of Fig.~\ref{fig:spiral_prop_6panel} shows $\Delta\text{time}$ vs the gas fraction of the disk, $f_{\text{gas}}$. 
There is not a strong relationship between $\Delta\text{time}$ and $f_{\text{gas}}$, and we find long-lived spirals in galaxies with both high (Remus) and low (m12b) $f_{\text{gas}}$.

It is fascinating that we find many instances of $m=2$ spiral episodes persisting for multiple Gyr.  
Moreover, we note that some separately identified spiral episodes could be part of a longer-lived spiral mode that may have faded in power for some time before recurring again as a dominant mode.
In Fig.~\ref{fig:all_sims_identified_spirals}, we see that many spiral amplitudes recur at similar $\Omega_p$.
For instance, there are two distinct episodes in m12f at $\sim$20~km/s/kpc that may be part of the same underlying spiral mode (green and light blue).
This is seen again in m12c, where there are recurring amplitudes at $\sim10$ and $\sim25$~km/s/kpc but the identified episodes are separated by time baselines where the spiral is not as strong. 
This suggests our disks may favor and host certain dominant spiral frequencies. 

\begin{figure}
\includegraphics[width=\linewidth]{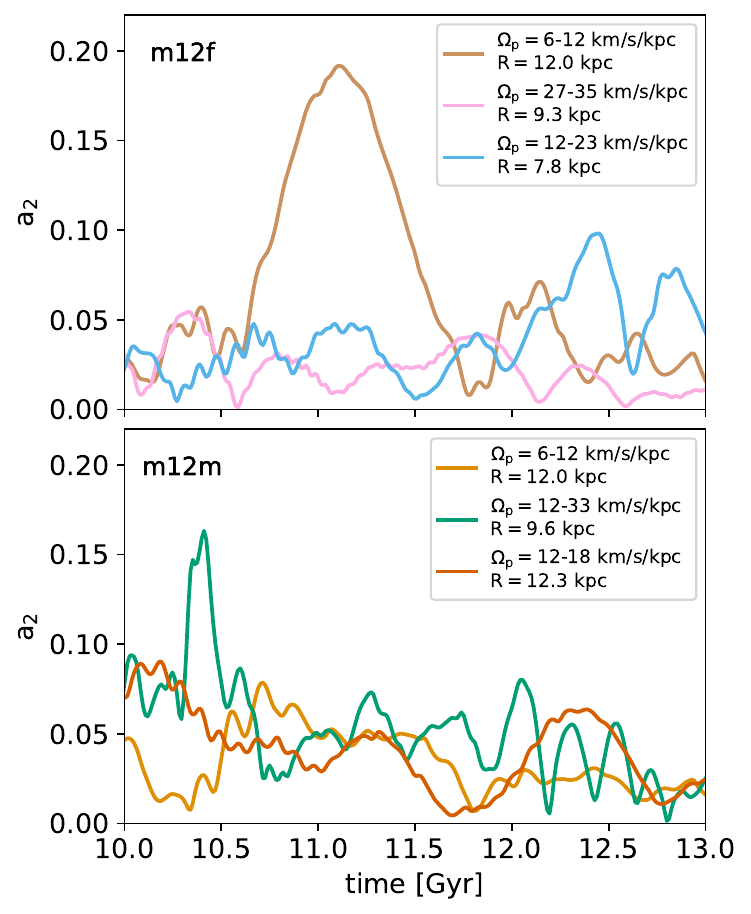}
\caption{Amplitude variation of three identified $m=2$ spiral episodes in m12f (top) and m12m (bottom).
Colors correspond to episodes of the same color in Fig.~\ref{fig:all_sims_identified_spirals}.
The amplitudes are obtained by performing an inverse Fourier Transform, applying a bandpass across the minimum and maximum frequency of the pattern $\pm \Delta\Omega$~km/s/kpc.
The amplitudes are calculated at $R_{\text{peak}}$, the radius of $a_{2,\text{max}}$.
Spiral episodes persist on Gyr timescales, but their amplitudes fluctuate on shorter timescales and exhibit oscillatory behavior.
We can isolate the range of radii and frequencies most impacted by tidal interactions.
For example, in m12f, the impact of the pericentric passage is strikingly clear in the slowest spiral (light brown), but not faster spirals (pink and blue). 
\label{fig:amplitude_variation}}
\end{figure}

Fig.~\ref{fig:amplitude_variation} shows the amplitude variation of three spiral episodes across $\sim$3~Gyr for galaxies m12f (top panel) and m12m (top panel). 
These spiral episodes exist on Gyr timescales, but their amplitudes fluctuate on much shorter timescales. 
The process for calculating these amplitudes over time is detailed in Appendix~\ref{app:inverse}. 
The amplitudes are reconstructed over the frequency range of the spiral $\pm \Delta\Omega \sim 2.1$~km/s/kpc at the radius of $a_{2,\text{max}}$.
The spiral amplitudes in Fig.~\ref{fig:amplitude_variation} are shown as the same colors used in their identification (Fig.~\ref{fig:all_sims_identified_spirals}).
This oscillatory nature is similar to what is observed in \cite{roskar2012} (their Fig.~6), where they identify long-lived spiral modes that fluctuate on shorter timescales.

The top panel of Fig.~\ref{fig:amplitude_variation} shows the amplitude evolution in m12f.
The light brown amplitude shows the most dominant spiral frequency that is likely excited via a pericentric passage of a satellite galaxy.
This is clearly displayed when the spiral amplitude drastically increases and peaks around the time of the interaction.
Afterwards, the amplitude dips to nearly zero before increasing to $\sim$0.06 within a short time span of $\sim$0.5 Gyr.
This spiral amplitude decreases in strength notably again in the next Gyr.
We see a similar oscillatory behavior in the other two amplitudes shown.
However, the light blue spiral amplitude oscillates on slightly longer timescales than the pink spiral amplitude, likely because of the slower $\Omega_p$ of the light blue amplitude.
The impact of the satellite passage cannot be clearly observed in the spiral frequencies shown in light blue or pink.

The bottom panel of Fig.~\ref{fig:amplitude_variation} shows the amplitude evolution in m12m.
The most dominant amplitude is shown in green and spans a large range of $\Omega_p$ as it increases in frequency over its lifetime.
This range of $\Omega_p$ overlaps completely with the spiral episode shown in orange-red, but the amplitudes are calculated at different radii.
There is a very weak correlation in the oscillatory behavior of these two spiral amplitudes as they decrease between $\sim10.5-11.0$~Gyr and then slightly increase in strength.
The fluctuation in strength is more apparent in the orange-red amplitude because of the smaller range of $\Omega_p$ being probed.
The spiral episode shown in light orange spans a slower range of $\Omega_p$, but at relatively large radii.
The larger oscillations for this amplitude occur over longer timescales likely because of the large radii and slow frequency.
Still, we see this spiral reaches a minimum amplitude around 10.3~Gyr and increases significantly over $\sim$0.3 Gyr before slowly decreasing again.

All of the spiral amplitudes in our simulations oscillate on relatively short timescales compared to how long they live.
While the fluctuations in strength are not particularly large, we focus on the fact that the changes are fractionally large relative to the amplitude range of these spirals.

\begin{deluxetable*}{cccccccccccc}
\tablecaption{Statistics of identified $m=2$ spiral episodes across the 10 FIRE-2 galaxies.\label{tab:spiral-stats}}
\tablehead{
  \colhead{Simulation} & \colhead{Instances} & \colhead{Instances/Gyr} & \multicolumn{3}{c}{$\Omega_p$} & \colhead{$\Delta$time} & \colhead{$\Delta R$} & \colhead{spiral disk}& \multicolumn{3}{c}{$a_{2\text{,max}}$}\vspace{-0.1cm} \\
   &  &  [Gyr$^{-1}$] & \multicolumn{3}{c}{[km/s/kpc]} & [Gyr] & [kpc] & fraction & \multicolumn{3}{c}}
\startdata
 &  &  & min & max & median & median & median & median & min & max & median \\ \hline
m12i & 6 & 1.54 & 10.5 & 31.4 & 19.9 & 1.25 & 5.0 & 0.6 & 0.030 & 0.069 & 0.050 \\
m12f & 9 & 1.56 & 4.2 & 33.5 & 18.8 & 1.50 & 7.3 & 0.8 & 0.033 & 0.192 & 0.070 \\
m12m & 4 & 0.88 & 8.4 & 31.4 & 19.4 & 2.25 & 9.2 & 0.8 & 0.049 & 0.170 & 0.071 \\
m12b & 6 & 1.15 & 6.3 & 54.5 & 28.8 & 1.63 & 5.0 & 0.9 & 0.041 & 0.168 & 0.091 \\
m12c & 8 & 1.80 & 8.4 & 35.6 & 18.8 & 1.50 & 6.2 & 0.8 & 0.039 & 0.096 & 0.067 \\
Romeo & 8 & 1.10 & 4.2 & 56.5 & 17.3 & 2.25 & 5.4 & 0.5 & 0.018 & 0.113 & 0.052 \\
Juliet & 9 & 1.75 & 8.4 & 48.2 & 28.3 & 1.25 & 3.3 & 0.4 & 0.019 & 0.058 & 0.028 \\
Romulus & 6 & 1.06 & 10.5 & 37.7 & 20.9 & 2.38 & 8.1 & 0.7 & 0.033 & 0.091 & 0.068 \\
Remus & 6 & 0.90 & 4.2 & 56.5 & 24.1 & 2.00 & 5.9 & 0.7 & 0.021 & 0.066 & 0.044 \\
Thelma & 6 & 1.81 & 6.3 & 33.5 & 18.8 & 1.25 & 6.3 & 0.6 & 0.022 & 0.125 & 0.040 \\ \hline
mean & 7 & 1.36 & 7.1 & 41.9 & 21.5 & 1.73 & 6.2 & 0.7 & 0.030 & 0.115 & 0.058 \\
median & 6 & 1.35 & 7.3 & 36.7 & 19.6 & 1.56 & 6.0 & 0.7 & 0.031 & 0.105 & 0.059 \\
\enddata
\tablecomments{Columns: (1) Name of the galaxy.
(2) Total number of spiral episodes identified. 
(3) Number of spiral episodes divided by total time of analysis. 
The total time of analysis is calculated as $t_{B-S}+(1/2)(S\Delta t)$~Gyr, where $t_{\text{B-S}}$ is the bursty to steady lookback time and $S\Delta t=1.5$~Gyr is the length of the time baseline. 
(4-6) Minimum, maximum, and median pattern speeds of spiral episodes. 
(7) Median duration of spiral episodes. 
(8) Median radial extent of spiral episodes. 
(9) Median fraction of the disk occupied by spiral episodes. 
(10) Minimum maximum amplitude $a_{2,\text{max}}$.
(11) Maximum maximum amplitude $a_{2,\text{max}}$.
(12) Median maximum amplitude $a_{2,\text{max}}$. 
The bottom two rows show the mean and median across each column, respectively.}
\end{deluxetable*}

\subsection{Radial Extent of $m=2$ Spiral Episodes} \label{ssec:radial_extent}

Across all galaxies, we see a large variety in radial extent, $\Delta R$, of spiral episodes with an overall median of $\Delta R=7$~kpc (Table~\ref{tab:spiral-stats}; column 8). 
We determine $\Delta R$ by identifying the range of radii with power $>5\times10^{-5}$ on the spectrograms (Fig.~\ref{fig:spectra}). 
We also define spiral disk fraction (Table~\ref{tab:spiral-stats}; column 9) as the fraction of the disk occupied by the spiral episode.
This fraction is the ratio of the spiral episode radial extent to the total radius of the disk we consider, which cuts out the inner regions if the galaxy hosts a bar and anything beyond $R_{90}$.
We find that $m=2$ spiral episodes in the isolated galaxies span larger $\Delta R$ (6.5~kpc vs 6.0~kpc) and occupy slightly larger disk fractions (0.7 vs 0.6) than the LG pairs.
This is interesting given the LG-like pairs have earlier forming, larger disks.
Although the spiral episodes in the isolated galaxies average slightly shorter lifetimes than the LG-like pairs, their spirals are more globally present across the disk.

Spirals in FIRE-2 span quite a large range across the disk, with the median disk fraction $>0.5$, indicating that these are large-scale features of the disk (i.e., grand design-like).
We note that the WDFT methodology we utilize to characterize spiral structure strongly favors the identification of global, grand design spirals over flocculent spirals, as larger-scale structures will necessarily have a stronger presence across the disk as a whole.
Our simulations certainly host many flocculent-like spirals \citep{orr2023}, but we do not focus on these smaller, more localized features.
It is the ubiquitous presence of global spiral modes that we focus on and is indicative of what drives spirals in our simulations.

The top left panel of Fig.~\ref{fig:spiral_prop_6panel} shows the stellar mass of the galaxy enclosed within $R_{90}$, $M_{90,*}$, versus the radius of peak power of the spiral mode normalized by $R_{90}$, $R_{\text{peak}}/R_{90}$.
We remind the reader that the WDFT analysis is biased towards identifying $R_{\text{peak}}$ at larger radii.
With this in consideration, we find that most spiral episodes occupy regions further out in the disk, with most of the episodes having $R_{\text{peak}}/R_{90}>0.5$.
Only a small handful of spirals occupy exclusively inner regions of the disk.
While there is not a strong relationship between $M_{90,*}$ and $R_{\text{peak}}/R_{90}$, we note that we do not see galaxies with very high stellar mass host spiral modes with $R_{\text{peak}}/R_{90}<0.5$.

The bottom left panel of Fig.~\ref{fig:spiral_prop_6panel} shows $\Delta\text{time}$ vs $R_{\text{peak}}/R_{90}$.
While the median $\Delta\text{time}$ slightly increases with $R_{\text{peak}}/R_{90}$, the $\pm 1\sigma$ deviation is large so we cannot conclude that there is any physically significant correlation between $\Delta\text{time}$ and $R_{\text{peak}}/R_{90}$.
Spiral episodes with large $R_{\text{peak}}$ persist for varying durations, but episodes with small $R_{\text{peak}}$ are generally shorter-lived, with a few exceptions. 
This is consistent with the idea that spirals at small radii have faster $\Omega_p$ and wind up on shorter timescales than slower patterns at larger radii. 

Remarkably, we find that strong tidal interactions do not necessarily generate long-lived spiral modes.
Rather, spirals associated with strong tidal interactions are short-lived ($\Delta\text{time}_{\text{median}}=1.00$~Gyr), but generally have a higher amplitude than isolated spirals.
The short-lived nature of tidally generated spirals may be because there is nothing else to excite the mode; thus it decays following the interaction.

The bottom right panel of Fig.~\ref{fig:spiral_prop_6panel} shows $\Delta\text{time}$ vs spiral disk fraction (the fraction of the disk the spiral mode occupies radially).
The spiral episodes in the isolated galaxies generally occupy $>50\%$ of the disk as compared to spirals in the LG-like pairs, which occupy a wider range of disk fractions.
The $\Delta\text{time}_{\text{median}}$ is the same in both groups, but with slightly more variation in the LG-like pairs.
Longer-lived spiral episodes are generally found in the absence of strong tidal interactions and occupy disk fractions $>0.5$.
Strong tidal interactions are associated with shorter-lived spiral episodes with disk fractions $>0.5$.
Spirals that have small to moderate $\Delta\text{time}$ in the absence of strong tidal interactions span a wide range of disk fractions.

\subsection{Strength of $m=2$ Spiral Episodes ($a_{2\text{,max}}$)} \label{ssec:a2max}

To compare spiral arm strength, we consider the maximum amplitude, $a_{2\text{,max}}$, of spiral episodes.
While $a_{2\text{,max}}$ does not give the full picture of how amplitude varies over time, it is nonetheless a useful measure to compare between different spiral episodes.
We calculate $a_{2\text{,max}}$ for each spiral episode by performing an inverse Fourier Transform using a bandpass filter around the frequency range of $\Omega_{p,\substack{\text{max}\\ \text{min}}}\pm\Delta\Omega$, where $\Delta\Omega\sim2.1$~km/s/kpc is the frequency resolution.
This procedure is detailed in Appendix.~\ref{app:inverse}.

In this paper, we consider strong spiral episodes to be above a threshold of $a_{2,\text{max}}=0.1$. 
We find the lowest $a_{2\text{,max}}=0.018$ in Romeo, and highest in m12f with $a_{2\text{,max}}=0.192$.
The overall median is $a_{2\text{,max}}=0.059$, respectively (Table~\ref{tab:spiral-stats}; column 12). 
Thus, the typical spirals in our simulations are not very strong.
All galaxies have their minimum $a_{2\text{,max}}$ between $\sim0.020-0.050$.
The range of maximum $a_{2\text{,max}}$ is much wider, between $\sim0.060-0.200$.
We find increasingly fewer spiral episodes at larger $a_{2\text{,max}}$ and only 8 of our 68 spiral episodes reach $a_{2\text{,max}}>0.1$.

We find that the isolated galaxies have generally higher $a_{2\text{,max}}$ values compared to the LG-like pairs.
In m12f, the high $a_{2\text{,max}}$ is generated by a pericentric passage of a satellite galaxy and occurs at 11.11~Gyr, immediately following the pericenter at 10.8~Gyr. 
The two strong spiral episodes in m12b also occur following a pericentric passage and prior to the merger event with the satellite galaxy, demonstrating that strong tidal interactions can lead to stronger spirals. 
Strong spiral structure can also emerge in isolation, as that this is the case for episodes with $a_{2\text{,max}}>0.1$ in m12m, Romeo, and Thelma.
The strongest spiral episode in m12m reaches its maximum amplitude when the galaxy is experiencing bursty star formation near the beginning of our analysis.
Increased star formation during this bursty phase likely occurred along spiral arms, leading to the increase in relative amplitude.

The overall smaller $a_{2\text{,max}}$ in the LG-like pairs may be due in part to the lack of strong tidal influence over the time of analysis.
Juliet has the lowest median $a_{2\text{,max}}=0.028$, is the least massive and physically smallest galaxy in our sample, and has a relatively high velocity dispersion even at present day \citep{mccluskey2023}.
The LG-like pairs are generally kinematically cooler than the isolated galaxies, which may result in the weaker, but somewhat longer-lived spiral amplitudes we find.
While strong spiral amplitudes can form in isolation, we find that the strongest spiral amplitudes are generated as a result of significant tidal perturbations.
We note that most strong spirals modes with $a_{2,\text{max}}>0.1$ coincide with a large scale tidal interaction, although tidal interactions do not necessarily lead to stronger spirals.

The top right panel of Fig.~\ref{fig:spiral_prop_6panel} shows $a_{2,\text{max}}$ vs the spiral disk fraction.
Spiral episodes that undergo a strong tidal interaction nearly always have disk fractions $>0.5$, whereas those in isolation vary widely.
Moreover, a majority of spiral episodes in the isolated galaxies occupy a disk fraction $>0.5$, and there is a more even distribution among the LG-like pairs.
There are very few spirals with large $a_{2,\text{max}}$ and small spiral disk fraction, possibly because the nature of spiral structure excited by tidal interactions is more grand design in appearance \citep[e.g.,][]{toomre&toomre1972}.
Overall, spiral episodes occupy both small and large fractions of the disk, showing the diversity in phenomenology of spiral structure.
The spiral episodes in FIRE-2 galaxies are generally weak and instances with very strong amplitudes are associated with tidal interactions.

\subsection{Stellar Age Cuts and Star Forming Gas} \label{ssec:star_forming_gas}

\begin{figure*}
\begin{center}
\includegraphics[width=0.9\linewidth]{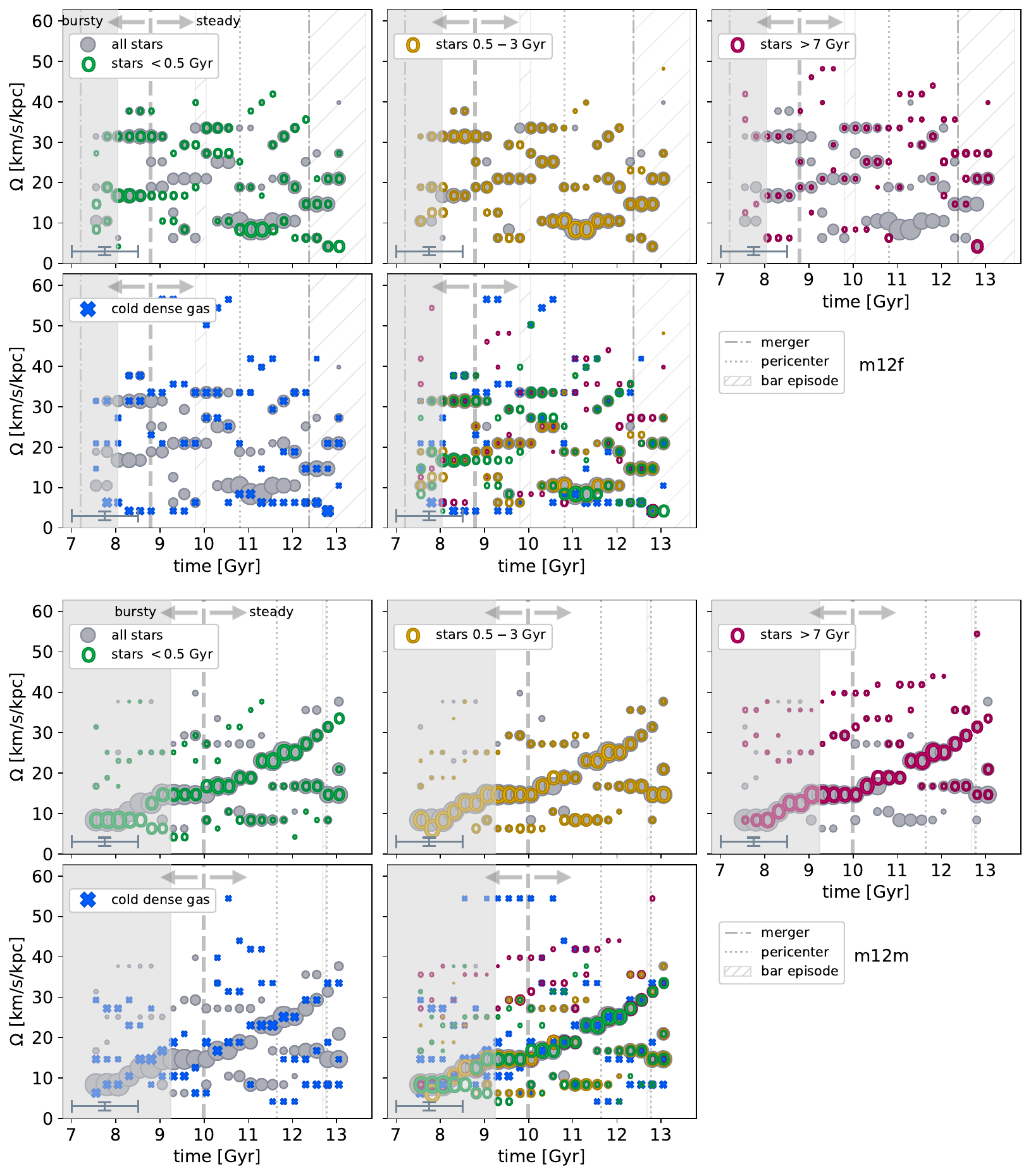}
\caption{Frequency evolution of the three most dominant $m=2$ amplitudes in stars with varying stellar ages and star forming gas ($T<10^4$ K and amu/cm$^3>10$) for m12f (top) and m12m (bottom).
Stars are divided into three different groups of stellar ages: $<0.5$~Gyr (green), $0.5-3$~Gyr (yellow), and $>7$~Gyr (magenta). 
The cold dense gas is shown in the blue crosses. 
The gray points show all stars and are the same as in Fig.~\ref{fig:all_sims_identified_spirals}.
The evolution of each group is shown in its own panel, as well as stacked on top of each other.
Points are sized proportional to their amplitudes in each time baseline.
The time baseline ($1.5$~Gyr) and frequency resolution ($\Delta\Omega_p\sim 2.1$~km/s/kpc) are shown in the bottom left.
The thick gray vertical dashed line indicates when the galaxy transitions from bursty to steady star formation rate.
The grayed out region shows the time period that we do not analyze the results for the galaxy.
We indicate when the galaxy undergoes merger events (dash-dot line), pericentric passages (dotted line), and bar episodes (hatched rectangle).
In m12f, stars $<0.5$~Gyr and $0.5-3$~Gyr trace the pattern speeds of all stars more strongly than stars $>7$~Gyr, particularly in the spiral amplitude that emerges after the pericentric passage at $\sim11$~Gyr.
All stellar populations in m12m closely follow the most dominant spiral pattern, and there is more overall agreement between the different age groups.
The cold dense gas in both m12f and m12m match well the pattern speed of the stars in many cases, but include instances where the two populations deviate from each other.
\label{fig:bubble_stars_gas_conjoint_m12f_m12m}}
\end{center}
\end{figure*}

We here analyze the frequency evolution of the strongest amplitude $m=2$ modes in three stellar age populations as well as cold dense gas ($T<10^4$ K and amu/cm$^3>10$).
Fig.~\ref{fig:bubble_stars_gas_conjoint_m12f_m12m} illustrates this time evolution for m12f (top set) and m12m (bottom set).
The time evolution for all stellar populations combined is shown in gray, where stellar populations are defined by their age at the time of analysis: $<0.5$~Gyr (green, youngest), $0.5-3$~Gyr (yellow, intermediate), and $>7$~Gyr (magenta, oldest), and cold dense gas (blue cross).

We note that analyzing stars younger than 0.5~Gyr over a 1.5~Gyr window of time may result in the detection of potentially spurious amplitudes as the population of stars analyzed changes over the window of time.
However, if we expect star forming gas to trace strong spiral arms, the resulting evolution of the youngest stars should also trace the same spiral mode as all stars.
And indeed, we do see in Fig.~\ref{fig:bubble_stars_gas_conjoint_m12f_m12m} that the youngest stars generally trace the evolution of all stars well.

In m12f, the strongest modes for all stars combined are driven by the young and intermediate population.
This is apparent when considering that the dominant modes for stars $<0.5$~Gyr and $0.5-3$~Gyr closely follow the frequencies and amplitudes of all stars combined.
Stars $<0.5$~Gyr broadly follow the pattern speeds of all stars but there is some deviation around the time of the pericentric passage for $\Omega_p\sim10$~km/s/kpc.
This population of very young stars is kinematically cold and thus more responsive to tidal interactions.
However, the power spectrum of stars $<0.5$~Gyr does not have a sharp peak so the assigned pattern speed's value is somewhat uncertain.

In some cases, the highest amplitude modes of a given population do not correspond to the highest amplitude modes for all stars combined.
This is particularly pronounced in stars $>$7~Gyr in m12f.
This is presumably associated with a higher velocity dispersion for these populations.  
For example, this oldest stellar population does not exhibit any dominant signature induced from the pericentric passage.

The lower, right hand panels of Figure~\ref{fig:bubble_stars_gas_conjoint_m12f_m12m} overlays all the dominant patterns for all populations considered. 
In some cases, the contributing populations to the evolution of the spiral pattern changes with time.
An example occurs in m12f $\sim10$~Gyr with an initial $\Omega_p\sim35$~km/s/kpc. 
The contributing population starts in stars $>7$~Gyr and cold dense gas.
The next few time baselines show that all stellar populations and cold dense gas follow along this dominant amplitude before it becomes only detectable again in old stars and cold dense gas.

The evolution of cold dense gas is more difficult to assess due to the complexity of its structure.
Particularly, we note that a low frequency structure is identified in many of our simulations, which may be a spurious signal arising from noise present in the data.
However, we find that the cold dense gas follows the same or similar pattern speeds as stars in the majority of instances.

The bottom set of panels in Fig.~\ref{fig:bubble_stars_gas_conjoint_m12f_m12m} shows the same evolution of stellar populations and cold dense gas in m12m. 
The evolution of spiral amplitudes in the different age populations is very markedly different from m12f.
We see remarkable agreement between all stellar populations in the evolution of pattern speed and relative amplitude of the most dominant spiral mode.

However, there are instances when the age groups do not match each other.
From around $11-12$~Gyr, there is a spiral amplitude around $10$~km/s/kpc that is dominated by the young and intermediate age stars.
During the same time period we see a dominant frequency at $\Omega_p\sim43$~km/s/kpc in the old stars.
There is also an interesting difference in pattern speeds from $10-11$~Gyr in stars $0.5-3$~Gyr and $\Omega_p\sim28$~km/s/kpc and stars $>7$~Gyr and $\Omega_p\sim35$~km/s/kpc.
The oldest stars appear to branch off from the intermediate stars around the time the galaxy transitions to steady star formation and there is some signature of stars $<0.5$~Gyr and cold dense gas around $30$~km/s/kpc during this time as well. 
The cold dense gas in m12m strongly follows the evolution of the most dominant spiral amplitude.

Whether or not all stellar age populations follow the same pattern speeds is critical in informing the underlying nature of spiral structure in our simulations.
Young stars are common tracers of spiral arms, but it is more difficult to discern if all stellar populations follow the same spiral patterns as each other.
If all stars follow along the same spiral amplitude, it suggests an underlying density wave-like feature present in the disk.
If only young stellar populations are found, the spiral amplitude may be a more material-like feature.

\begin{figure*}
\begin{center}
\includegraphics[width=0.9\linewidth]{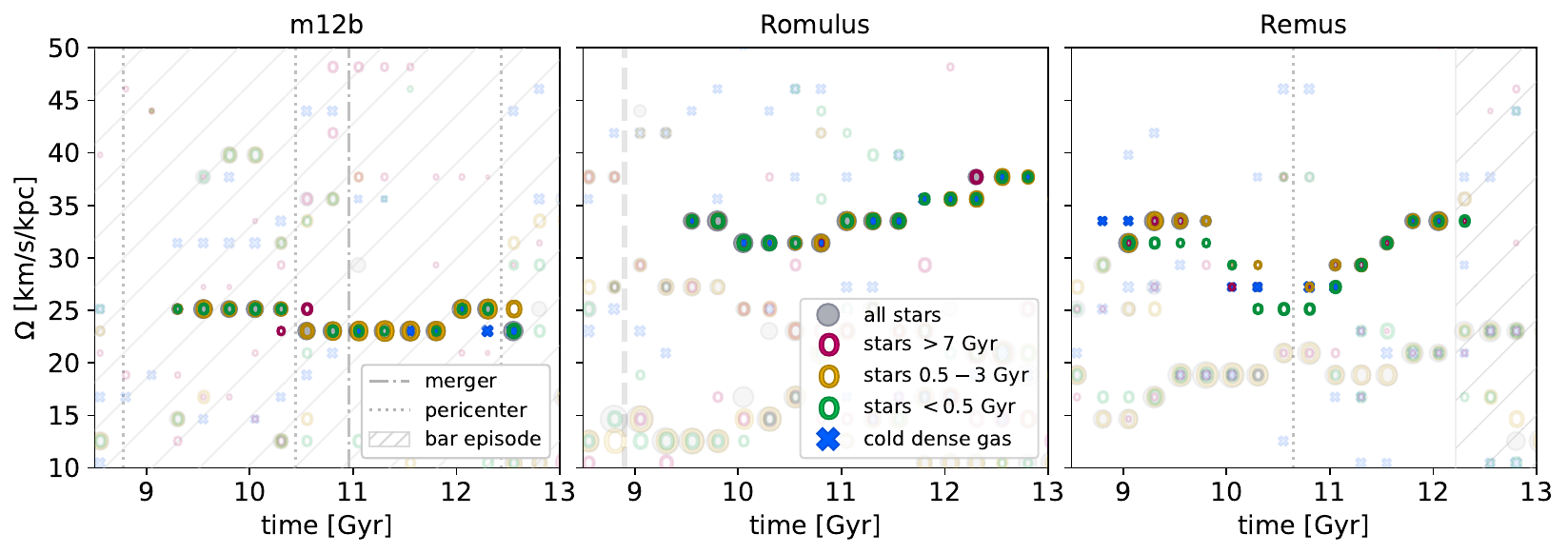}
\caption{Similar to Fig.~\ref{fig:bubble_stars_gas_conjoint_m12f_m12m}, but showing only the stacked plot and zoomed in on highlighted features of interest for m12b (left), Romulus (middle), and Remus (right).
In m12b, the spiral amplitude we focus on is only present in stars $<0.5$~Gyr and $0.5-3$~Gyr and is generated by a process that does not strongly impact all age groups.
Romulus shows an example of all populations of stars and cold dense gas following the same evolution of $\Omega_p$ and is generated via internal evolution of the disk.
Remus displays a spiral amplitude that we are only able to see the evolution of $\Omega_p$ clearly when considering the very young stars $<0.5$~Gyr. 
\label{fig:stars_gas_highlights}}
\end{center}
\end{figure*}

Fig.~\ref{fig:stars_gas_highlights} shows the overlaid evolution of frequencies for varying stellar ages and cold dense gas zoomed in on regions of interest for m12b, Romulus, and Remus.
This figure highlights some interesting features we find in the evolution of $m=2$ spiral amplitudes in varying populations of stars and cold dense gas.

The left panel of Fig.~\ref{fig:stars_gas_highlights} shows m12b, and we focus on the spiral amplitude with $\Omega_p\sim25$~km/s/kpc.
Stars $<0.5$~Gyr (green) and $0.5-3$~Gyr (yellow) dominate the evolution of this amplitude, and there is no strong signature in stars $>7$~Gyr (magenta).
The cold dense gas is also present in some, but not all, time baselines shown here.
m12b undergoes two pericentric passages and one merger event over this period of time, and the spiral structure excited by these tidal perturbations are most strongly associated with younger populations of stars.

The middle panel of Fig.~\ref{fig:stars_gas_highlights} shows Romulus, zooming in on a spiral amplitude at somewhat high pattern speeds, $\Omega_p\sim30-35$~km/s/kpc.
This spiral amplitude is present across all stellar populations and cold dense gas, suggesting that all stars are affected by and follow the same underlying mechanism that drives the presence and persistence of this feature.
The pattern speed across all populations matches even as $\Omega_p$ changes slightly over time.
This particular spiral episode is not very strong ($a_{2,\text{max}}=0.033$) but spans across a large radial range ($\Delta R=8.1$~kpc), which suggests that this spiral amplitude may be more density wave-like in nature.
This is particularly interesting considering all values of $\Omega_p$ associated with this spiral are relatively fast.

The right panel of Fig.~\ref{fig:stars_gas_highlights} shows Remus, and we zoom in on a spiral amplitude that begins $\sim33$~km/s/kpc, decreases to $\sim25$~km/s/kpc, and then speeds up to $\sim33$~km/s/kpc.
This spiral frequency changes significantly over time, but is interestingly only clearly present in the youngest stars $<0.5$~Gyr.
There is some signature of evolving $\Omega_p$ in stars $0.5-3$~Gyr and $>7$~Gyr but is not present throughout the evolution.
This suggests that the intermediate and older population of stars are impacted by what is causing $\Omega_p$ to slow down, but because they are kinematically hotter, they do not respond as strongly as the youngest stellar population.
The pericentric passage that occurs at $\sim$10.8~Gyr may be what is affecting the pattern speed of young stars, but we note that this satellite galaxy is not massive enough to meaningfully perturb the disk.

By analyzing different stellar ages, we are able to uncover the evolution of spirals that would not be apparent otherwise. 
While the contribution from different age populations vary for any given spiral episode, it is clear that coherently evolving spirals are found in all age populations across all galaxies we analyze (see Appendix~\ref{app:stars_cdgas}).
This strongly suggests that our spirals are driven by something more akin to density waves in nature than a purely material feature.

\section{Discussion} \label{sec:disc}
Analyzing the properties and time-evolution of galactic spiral structure is difficult, particularly in cosmological simulations.
However, we are entering an era where this is possible given the resolution of cosmological simulations such as in FIRE-2.
To our knowledge, Auriga simulations \citep{grand2017, grand2024} are the only other cosmological zoom-in simulation with work characterizing galactic spiral structure.
We note that TNG50 \citep{pillepich2024}, a large volume cosmological simulation, has recently investigated spiral structure and is discussed later in this section.

\cite{grand2016b} analyzes one Auriga simulation with a strong two-armed grand design spiral to investigate how their spiral arms impact radial redistribution and affect the metallicity gradient. 
Their results are similar to what has been found previously in N-body simulations; that their spirals are transient, winding density enhancements, which we find to be similar in nature to the spirals in FIRE-2 simulations.
We note their analysis is done using young stars $<3$~Gyr.

Galaxies from the Auriga simulations \citep{grand2024} are shown to have strong grand design spiral arms dominated by the $m=2$ multiplicity from visual inspection of the face-on projected stellar density \citep[see Fig. 1 in][]{grand2016a}. 
\cite{grand2016a} also note they measure strong $m=1$ (lopsided) amplitudes and that the strength of all $m$ multiplicities sharply increases during perturbations from satellite galaxies, both of which we also find in FIRE-2 simulations.
Their dominant $m=2$ Fourier amplitudes range $\sim0.05-0.1$, which are slightly higher than in our simulations ($\sim0.01-0.07$).
Our weaker amplitudes are likely because FIRE-2 galaxies exhibit many flocculent-like features.
While our amplitudes are generally weak, they are still within the range of amplitudes found in more idealized simulations \citep[e.g.,][]{roskar2012, baba2013}.

\cite{grand2016a} investigate the sources of vertical heating in Auriga simulations and conclude that the dominant contributor is not from spiral arms, but rather bars.
\cite{mccluskey2023} suggest that in FIRE-2 simulations, spiral arms likely drive overall heating in young stars, although we note this includes both radial and vertical heating. 
Additionally, FIRE-2 bars are not strong \citep{ansar2025}, so they are less likely to be a primary source of vertical disk heating. 
We defer exploring the impact of dynamical heating from spiral arms to future work.

Differences in the implementation of numerical models between FIRE-2 and Auriga may explain some of the variations in the properties of the resulting galaxies.
Auriga implements a pressurized, two-phase ISM compared to FIRE-2's multi-phase ISM, and the specific stellar feedback models in either simulation groups differ. 
As a result, Auriga galaxies experience less clustered star formation, leading to the formation of disks with fewer flocculent features than FIRE-2 galaxies. 

In the TNG50 large volume cosmological simulations, \cite{ghosh2025} characterize spiral features in 43 galaxies and finds that their disks host predominantly two-armed stellar spirals, with more multi-armed features in the gas, consistent with our results.
They find that colder stellar disks with higher gas fractions host stronger spirals, and while the former is true in FIRE-2 galaxies, we do not find a relationship between gas fraction and spiral amplitude. 
This discrepancy may be due to simulation differences (large volume vs zoom-in) or as a result of their larger sample size (43 vs 10).

Spiral arms in N-body simulations have consistently been shown to be transient features \citep[e.g.,][]{wada2011, grand2012b, sellwood2011}, and we find this to be the case in FIRE-2 galaxies as well.
However, while individual spiral arms may be transient in the sense that they are continuously forming and breaking apart, the underlying spiral modes can last on Gyr timescales, with spiral amplitudes tending to recur at similar frequencies.
Indeed, \cite{roskar2012} show that the evolution of their dominant spiral modes persist for $\sim2$~Gyr, which is comparable to the $\Delta\text{time}$ of our spiral episodes (see Fig.~\ref{fig:spiral_prop_6panel}).

Moreover, the transient nature of spiral arms in N-body simulations has been demonstrated across a range of varying galactic properties.
For example, the galaxy analyzed in \cite{grand2012a} does not have one single dominant $m$ mode, but rather has similar amplitudes across the $m=2-4$ multiplicities.
Still, their spiral arms co-rotate with the disk at nearly all radii and serve as significant drivers of radial migration.
While the spirals in our simulations are not co-rotating, as our identified episodes span multiple kpc in radii, they generally exist between the ILR and OLR with decreasing $\Omega_p$ as a function of radius.

\cite{fujii2011} utilizes high resolution stellar disks and find their Fourier amplitudes are generally $<0.1$ with many multiplicities reaching $<0.05$ over the disk's evolution, which is similar to the range of amplitudes we calculate.
Their dominant Fourier modes are $m=4-6$ compared to $m=2$ in our simulations, though some of our galaxies display $m=3-4$ strengths comparable to $m=1$ and $m=2$ during some points in their evolution.

\cite{wada2011} utilize N-body + SPH simulations to investigate how gas spirals evolve. 
Their disks evolve many transient spiral arms, persisting on the scale of 100 Myr.
Both star and gas particles follow the galactic rotation curve in their simulations, and gas particles fall into stellar spiral potentials from either side of the spiral arm.
We similarly find the pattern speed of gas spirals in our simulations generally follow that of stellar spirals (see \S~\ref{ssec:star_forming_gas}).
\cite{wada2011} note that the pattern speed of the gas and stars may differ from each other in cases such as during a tidal interaction, and we do find instances where $\Omega_p$ of stars and gas deviate during such events.

\cite{grand2012b} investigate the impact of a bar on the behavior of spiral arms using an N-body + SPH simulation of a barred spiral galaxy. 
The overall dynamics of spiral arms is similar to what is presented in \cite{grand2012a}; however, they find the presence of a strong bar may boost $\Omega_p$ to be slightly faster than the galactic rotational velocity.
The strength of the bar significantly decreases at later times in their simulation, and $\Omega_p$ also decreases to follow the rotation curve at all radii.

Similarly, \cite{roca-fabrega2013} utilize simulations of barred and unbarred galaxies to identify the differences in the resulting $\Omega_p$. 
In their strongly barred galaxy, they find that $\Omega_p$ of the $m=2$ spiral is nearly constant across radius.
For the unbarred case, their results agree with previous works confirming that the spirals follow the galactic rotation curve.
They note that as the bar strength weakens ($A_2/A_0 < 0.1$), the spiral arms begin to follow the rotation curve rather than propagate as a solid body.
We do not find any instance in our simulations where the bar is strong enough to excite a single pattern speed that mimics solid body rotation.

As with \cite{grand2012a}, we find that our $m=2$ spiral modes do not strictly follow the galactic rotation curve.
Rather, the overall dynamics of our spiral amplitudes loosely follow that of the rotation curve, but the radial ranges of each spiral pattern is present over a range of radii.
For galaxies hosting a bar, we find that $\Omega_p$ of the dominant spiral modes do not seem to be directly correlated with the pattern speed of the bar, calculated in \cite{ansar2025}.
However, we do find instances where $\Omega_p$ increases prior to the emergence of a bar episode.
In these cases, the bar may be acting to boost $\Omega_p$, or the spiral may be driving the formation of the bar.
Additionally, some very fast spiral patterns we identify are likely due to the bar.
While we cannot directly resolve the pattern speed of the bar, measurements from \cite{ansar2025} show that the bar rotates at much faster speeds than what we find for the dominant spiral modes and are above the Nyquist Frequency.

\cite{kumar2021} and \cite{kumar2022} assess the impact of tidal fly-bys in N-body simulations and find these interactions produce strong $m=2$ spiral arms that live for multiple Gyrs.
With the inclusion of gas dynamics, \cite{pettitt2016} and \cite{pettitt&wadsley2018} find that tidally generated spirals in their simulations are also predominantly two-armed structures that are transient, but persist on Gyr timescales. 
They calculate $\Omega_p$ of the induced spiral to lie somewhere between co-rotation and the ILR, further supporting the findings of previous works \citep[e.g.,][]{oh2008, dobbs2010}.
These results are in agreement with what we see in FIRE-2 simulations; however, we note that the spiral structure resulting from satellite interactions is largely dependent on the properties of both the host and companion galaxy \citep[e.g., mass ratio, direction of orbit, as explored in,][]{pettitt&wadsley2018, kumar2022}.
Given the diversity of satellite interactions across FIRE-2 galaxies, we reserve a more detailed comparison and investigation of the impact of tidal perturbations on spiral structure to future work.

\cite{baba2013} utilize N-body simulations to analyze stellar dynamics during the growing and damping periods of spiral arms. 
The amplitudes of their spiral arms grow as their disk evolves, but generally remains $<0.1$ and all modes contribute to the spiral structure in the disk.
They find the spiral arms in their simulations can be explained via swing amplification and that $\Omega_p$ at each radius generally follows the rotation curve. 
They note their spiral arms appear `quasi-steady' on the global scale, but local spirals are not `static', which is what we find in FIRE-2 simulations.

\cite{khoperskov2018} and \cite{debattista2025} analyze azimuthal metallicity variations in isolated simulations and find spirals present in all age populations, but are strongest in younger populations.
\cite{ghosh2025} also find spiral features in all stellar ages in TNG50 simulations.
Furthermore, \cite{ardevol2025} characterizes spiral arms across different age populations in N-body + SPH simulations using a newly developed method that measures the `local dimension' of the stellar density across the disk.
They find the presence of spiral arms in all stellar populations and describe differences in spiral properties across age groups.
This is in agreement to what we see in FIRE-2 galaxies; stars of all ages and star-forming gas are often present in dominant spirals, but the younger, kinematically cooler stars are most responsive to spiral perturbations.

Differentiating the behavior of our spirals of different ages may provide insight into the spirals of the MW and external galaxies. 
Observations of both the MW \citep[e.g.,][]{gaia2023, uppal2023} and of external galaxies \citep[e.g.,][]{shabani2018, yu&ho2018, peterken2019} find spiral structure that display different morphologies across different populations.
We have not measured in detail how young and old spirals differ in this paper, but emphasize that we do detect spirals in the separate populations (see Fig.~\ref{fig:bubble_stars_gas_conjoint_m12f_m12m}) and have established a foundation for further work investigating spirals in FIRE-2 galaxies. 
The presence of spirals in all age populations is a key result for considering the mechanisms that drive their formation and persistence.
More importantly, these results are consistent with what is seen in observations, and further detailed studies can help us better understand the spirals in our Universe.

\section{Conclusion} \label{sec:conc}
In this paper, we analyze the properties, characteristics, and evolution of dominant $m=2$ spiral amplitudes in ten FIRE-2 simulations and present a catalog of our results.
We perform this analysis for the first time in cosmological simulations and on stars of varying stellar ages and star forming gas.
We have demonstrated, in FIRE-2 simulations, that galactic environment and merger history have significant influence on the resulting spiral structure and that there are multiple processes that stimulate spiral structure in our galaxies. 
Our key conclusions are:

\begin{itemize}
    \item All Milky Way-mass galaxies in the FIRE-2 suite of simulations have robust spiral structure throughout their history.  Each galaxy hosts multiple spiral patterns simultaneously with different $\Omega_p$. The dominant spiral $m=2$ amplitudes tend to recur at particular $\Omega_p$ over the evolution of a given disk.
    \item Spiral $m=2$ patterns often persist over Gyr timescales but fluctuate significantly in amplitude on an order magnitude shorter timescales (Fig.~\ref{fig:amplitude_variation}). The strongest spirals at a given time in FIRE-2 simulations are generally weak, for example, the maximum $m=2$ Fourier amplitude across simulations is $a_{2,{\rm max}}=0.192$ and the mean value for the median amplitude in a given galaxy is $a_{2,{\rm max}}=0.058$ (Fig.~\ref{fig:global_amplitudes}~\&~\ref{fig:spiral_prop_6panel} and Tbl.~\ref{tab:spiral-properties}~\&~\ref{tab:spiral-stats}).
    \item Strong tidal interactions generate spirals that radially span $>50$\% of the disk, but are remarkably short-lived (median duration 1.0~Gyr). Strong tidal forces do not always lead to high amplitude $a_{2,\text{max}}$, but we find the strongest spirals across our sample are those affected by tidal interactions (Fig.~\ref{fig:spiral_prop_6panel}).
    \item We find that spirals with very fast $\Omega_p$ are predominantly correlated with bar episodes. There are also instances where $m=2$ spiral episodes increase $\Omega_p$ leading up to a bar episode (Fig.~\ref{fig:all_sims_identified_spirals} and Fig.~\ref{fig:all_sims_bubble}).
    \item Galaxies with kinematically cooler disks (m12f, m12m, m12b, Romeo, Romulus), compared to relatively hotter disks (m12i, m12c, Juliet, Thelma), can host spirals that are longer-lived (mean duration 2.18~Gyr vs 1.53~Gyr), and stronger (mean $a_{2,\text{max}}$ 0.07 vs 0.05). The spirals in these kinematically cooler disks tend to span a slightly larger radial extent as well (mean disk fraction 0.7~kpc vs 0.6~kpc).
    \item We find that stars split into groups with varying stellar ages, as well as cold dense gas, generally follow the same evolution of $\Omega_p$. This suggests that our spirals are density wave driven features, as all stellar populations and gas are affected by the same underlying internal processes that drive dominant patterns. The instances that these stellar age groups differ in $\Omega_p$ occur mostly during tidal interactions or during a bar episode. When these age groups deviate, old stars generally have a more dominant signature at higher $\Omega_p$ associated with being closer towards the center of the disk, whereas younger stars are generally dominant at slower $\Omega_p$, further out in the disk. (Fig.~\ref{fig:bubble_stars_gas_conjoint_m12f_m12m} and Fig.~\ref{fig:all_sims_bubble_stars_cdgas}).

\end{itemize}

Our simulations display spiral structure with diverse properties, spanning a variety of pattern speeds, strengths, radial ranges, and lifetimes.
There are multiple spiral patterns present in a galaxy at any given moment in time, supporting the idea that spiral arms are transient, evolving features rather than a rigid density wave rotating at a constant pattern speed.
The diversity of spiral properties indicates that there are likely multiple mechanisms needed to explain the various evolution of structure that we see within our simulations.

This is the first paper in a series that will carefully investigate the nature of spiral structure in FIRE-2 simulations.
Future work will investigate the connection of spiral structure with breathing and bending waves, a deeper analysis of infalling satellites and bars, and how the implementation of different simulation physics impacts spiral structure. 

\begin{acknowledgments}
We thank the referee for their constructive feedback that greatly helped strengthen the paper.
SRL, KJD, $\&$ JRQ acknowledge support from NSF grant AST-2109234.
SRL $\&$ JRQ acknowledge support from NSF grant AST-2511388.
SRL additionally acknowledge HST grant AR-16624 from STScI.
AW received support from: NSF via CAREER award AST-2045928 and grant AST-2107772; NASA ATP grant 80NSSC20K0513; HST grant GO-16273 from STScI.

We generated simulations using: XSEDE, supported by NSF grant ACI-1548562; Blue Waters, supported by the NSF; Frontera allocations AST21010 and AST20016, supported by the NSF and TACC; Pleiades, via the NASA HEC program through the NAS Division at Ames Research Center.

KJD and LBeS acknowledge support from the Heising Simons Foundation grant \# 2022-3927. They also respectfully acknowledge that the University of Arizona is home to the O'odham and the Yaqui.
We respect and honor the ancestral caretakers of the land, from time immemorial until now, and into the future.

FIRE-2 simulations are publicly available \citep{wetzel2023} at \url{http://flathub.flatironinstitute.org/fire}.
Additional FIRE simulation data is available at \url{https://fire.northwestern.edu/data}.
A public version of the \textsc{Gizmo} code is available at \url{http://www.tapir.caltech.edu/~phopkins/Site/GIZMO.html}.
\end{acknowledgments}

\software{IPython \citep{ipython}, Matplotlib \citep{matplotlib}, Numpy \citep{numpy}, Scipy \citep{scipy}, \texttt{halo\_analysis} \citep{haloanalysis}, \texttt{gizmo\_analysis} \citep{gizmoanalysis}}

\bibliography{bibliography}{}
\bibliographystyle{aasjournalv7}

\appendix
\section{Global Fourier Amplitudes} \label{app:global_amplitude}

\begin{figure*}[h]
\begin{center}
\includegraphics[width=0.9\linewidth]{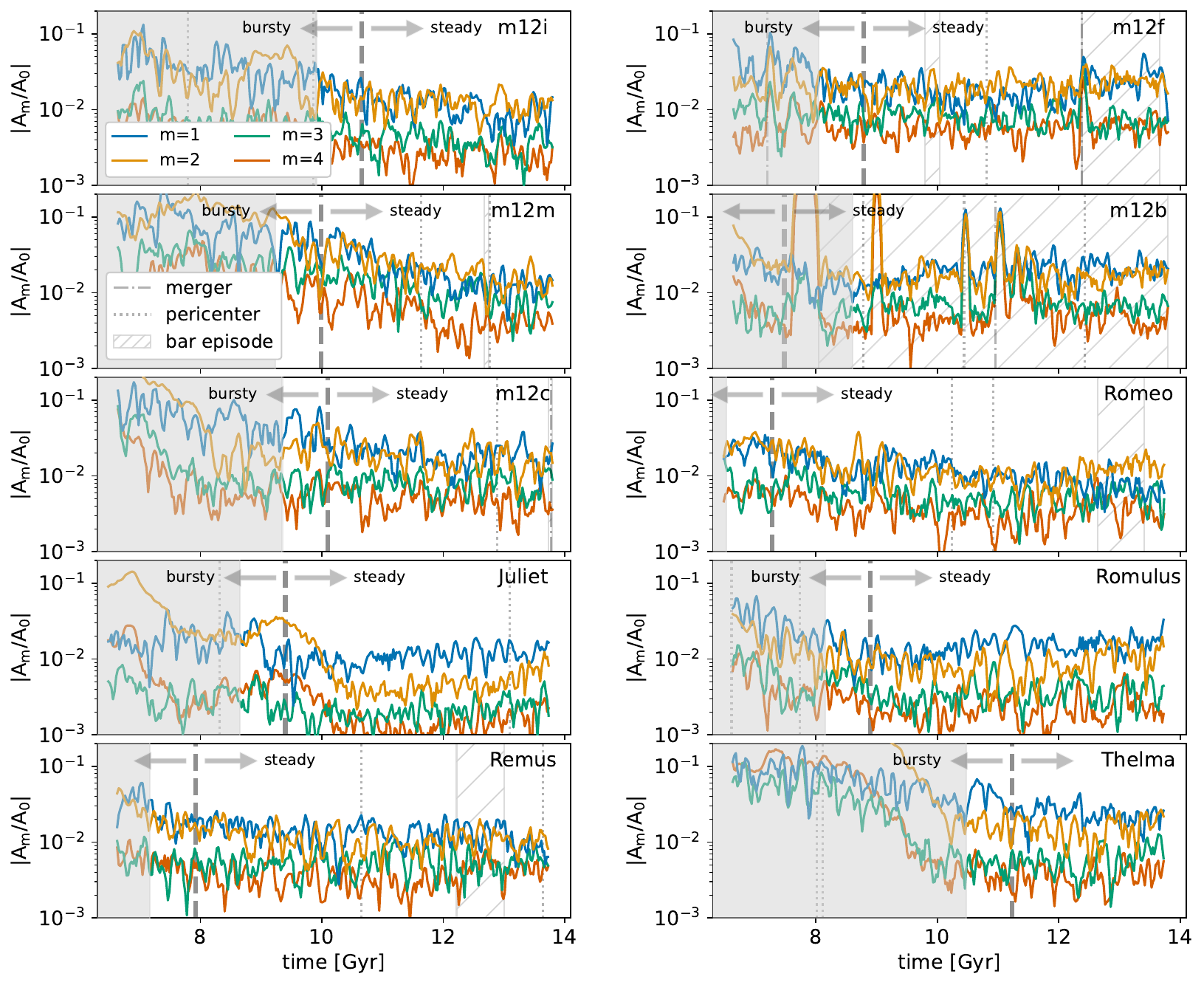}
\caption{Expanded version of Fig.~\ref{fig:global_amplitudes}.
Global Fourier amplitudes for the m=1-4 multiplicities shown across time for all FIRE-2 galaxies analyzed.
The thick gray vertical dashed line indicates when the galaxy transitions from bursty to steady star formation rate.
The grayed out region shows the time period that we do not analyze the galaxy.
We indicate when the galaxy undergoes merger events (dash-dot line), pericentric passages (dotted line), and bar episodes (hatched rectangle).
\label{fig:all_sims_global_fourier_amplitudes}}
\end{center}
\end{figure*}

In Fig.~\ref{fig:all_sims_global_fourier_amplitudes}, we present the evolution of $m=2$ Fourier amplitudes for all 10 FIRE-2 galaxies analyzed, calculated with Eq.~\ref{eq:2} using all star particles and normalized by $A_0$.
Each data point plotted is smoothed by applying a moving average over three snapshots for visual clarity. 
The $m=1$ and $m=2$ multiplicities are nearly always dominant across all simulations, and in most cases, the amplitude of both multiplicities are comparable to each other.
We consider the $m=2,3,4$ multiplicities for studying spiral structure and note that $m=2$ is stronger than $m=3,4$ for all simulations. 
A strong $m=1$ multiplicity indicates a lopsided disk, which is particularly strong in Juliet and Romulus.
m12f and m12b experience strong tidal interactions that largely affect all $m$-multiplicities shown.
In many galaxies, such as m12i, m12m, m12c, Juliet, and Thelma, we see that the $m=1,2$ multiplicities are much stronger before the galaxy transitions to steady star formation and decrease in strength thereafter.
The increased amplitude prior to transition is likely due to increased star formation along $m=2$ spiral arms, leading to higher overdensities along the arms.

\clearpage
\pagebreak

\section{Spectrograms} \label{app:spectrogram}

\begin{figure*}[hb]
\begin{center}
\includegraphics[width=0.8\linewidth]{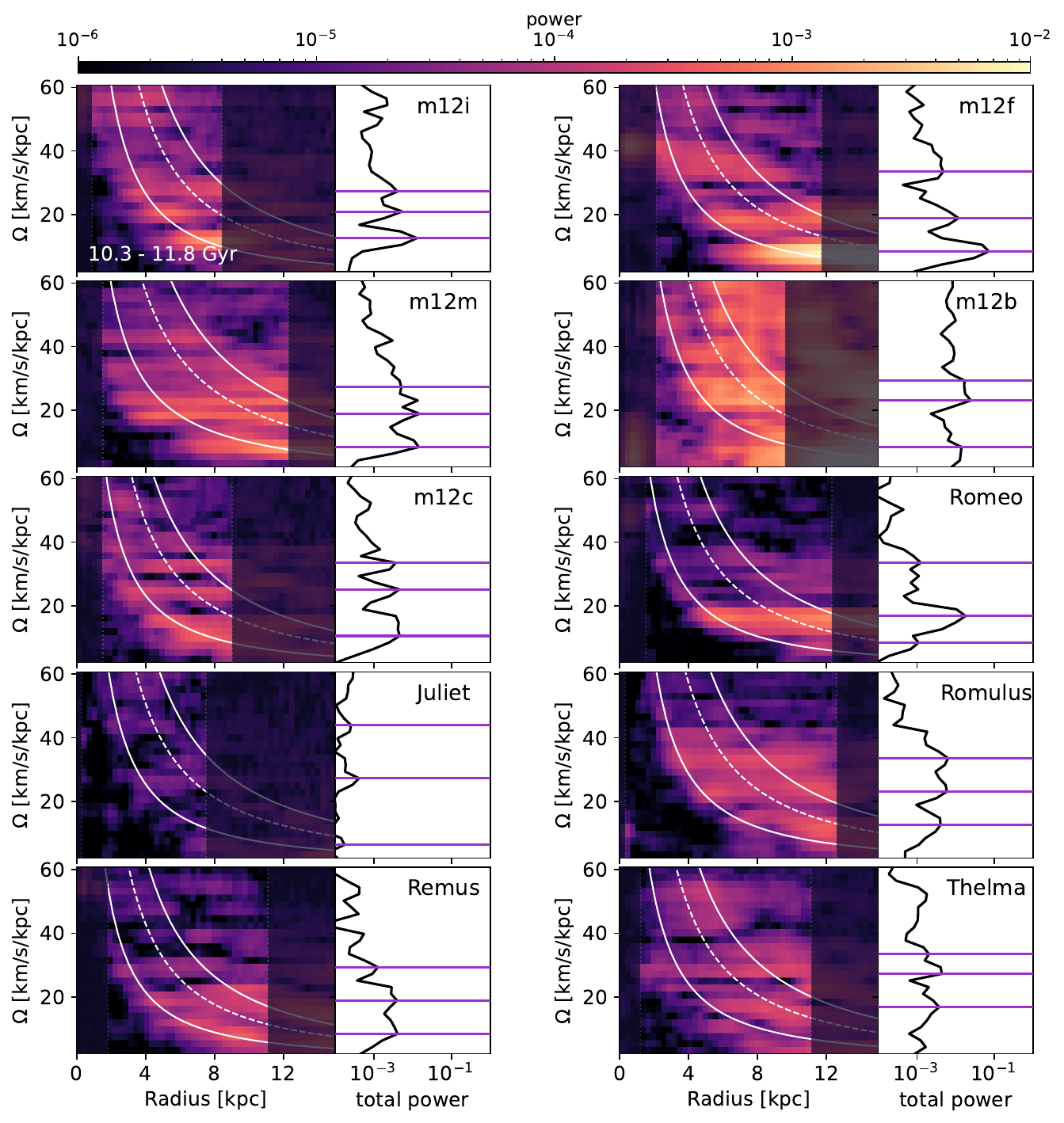}
\caption{Similar to Fig.~\ref{fig:spectra}.
Power spectra for all FIRE-2 galaxies analyzed over the $10.3-11.8$~Gyr time baseline.
The left side of each plot shows the spectrogram of pattern speeds as a function of radius (Eq.~\ref{eq:6}), colored by power, and the right side shows the radially integrated power (Eq.~\ref{eq:total_power}).
More power indicates stronger m=2 spiral structure present over the time baseline.
The darkened rectangles cut out the bar and region beyond the disk and are not included in the integrated spectra.
The Inner Linblad, Outer Linblad, and corotation resonances are plotted as the two solid white lines and dashed white line, respectively.
The purple horizontal lines show peaks in the power spectrum and identify the three most dominant pattern speeds. 
\label{fig:all_sims_spectra}}
\end{center}
\end{figure*}

Fig.~\ref{fig:all_sims_spectra} shows the spectrogram and power spectrum for all 10 FIRE-2 galaxies analyzed using the methodology of \S\ref{ssec:WDFT}.
There is diversity in how $m=2$ spiral structure is expressed across these galaxies.
The disk in m12b is perturbed over this period of time, resulting in a spectrogram with power spread over nearly all radii and $\Omega_p$.
Romeo has one particularly strong spiral pattern that spans a large radial extent.
In contrast, Juliet does not have any strong $m=2$ spiral structure evolve over the time baseline shown, and its disk is quite small at this point in time.
Across all galaxies, we see that power is generally concentrated between the ILR and OLR, though in some cases, such as m12m, the power extends beyond the ILR and OLR.

\clearpage
\pagebreak

\section{Time-evolution of Spiral Structure} \label{app:bubble}

\begin{figure*}[h]
\begin{center}
\includegraphics[width=0.75\linewidth]{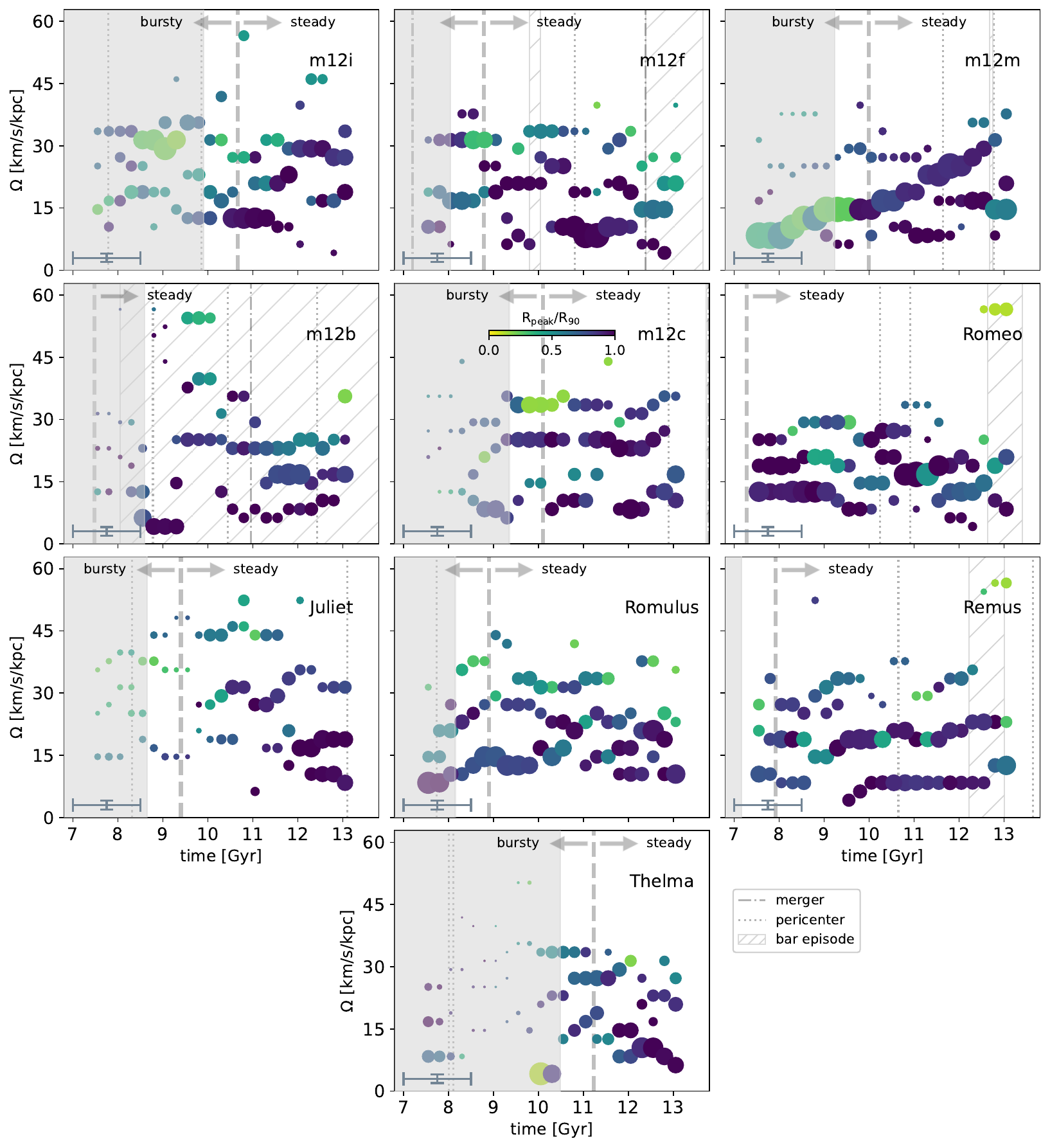}
\caption{Expanded version of Fig.~\ref{fig:bubble_m12f_m12m}.
Frequency evolution of the three most dominant m=2 amplitudes for all FIRE-2 galaxies analyzed.
Points are color-coded by the radius of peak spiral power relative to $R_{90}$, and the size is proportional to their amplitudes during the relevant time baseline. 
The time baseline (1.5~Gyr) and frequency resolution ($\Delta\Omega_p\sim 2.1$~km/s/kpc) are shown in the bottom left. 
The thick gray vertical dashed line indicates when the galaxy transitions from bursty to steady star formation rate.
The grayed out region shows the time period that we do not analyze the galaxy.
We indicate when the galaxy undergoes merger events (dash-dot line), pericentric passages (dotted line), and bar episodes (hatched rectangle).
\label{fig:all_sims_bubble}}
\end{center}
\end{figure*}

Fig.~\ref{fig:all_sims_bubble} shows the frequency evolution of the dominant $m=2$ spiral patterns for all 10 FIRE-2 galaxies analyzed.
An interesting feature present across most galaxies, though most notably in m12i, m12m, and Juliet, is that the relative location of peak power moves outwards in the disk once the galaxy transitions to steady star formation.
This may be due to star formation concentrated in the inner regions of the disk during the bursty phase, leading to stronger $m=2$ amplitudes at lower radii.
Once the galaxy transitions to steady star formation, there may be less $m=2$ overdensity at low radii, causing the relative location of power to move outwards.
m12i and Juliet in particular have physically small disks (see location of $R_{90}$ in Fig.~\ref{fig:all_sims_spectra}) and we can see the location of power moving outwards quite dramatically.

\clearpage
\pagebreak

\section{Identifying $m=2$ Spiral Episodes} \label{app:identify_arms}

Here, we describe the procedure developed to connect the dominant $m=2$ spiral amplitudes identified over time.
To begin, we consider the three most dominant features as identified by the purple horizontal lines in Fig.~\ref{fig:all_sims_spectra}.
We threshold the spectrograms (e.g., Fig. \ref{fig:all_sims_spectra}) to ensure that all $m=2$ spiral arms satisfy a minimum power cutoff of $5\times10^{-5}$ across the radii of analysis.
This threshold was selected as the minimum power cut that identifies all cohesively evolving spiral amplitudes.
Using a more stringent threshold fails to catch clear-cut instances of spiral episodes.

We then consider the time-evolution of the dominant $m=2$ frequencies, shown in Fig. \ref{fig:bubble_m12f_m12m}.
We only identify spirals at times adjacent to and beyond the bursty to steady transition $t_{\text{B-S}}$.
More specifically we begin analysis at the lookback time of $t_{\text{B-S}}+(1/2)(S\Delta t)$, where $S\Delta t$ is the length of the time baseline used.
This transition time \citep{yu2021} occurs generally $2$~Gyr after the galaxy becomes dominated by ordered rotation \citep{mccluskey2023}.
We decided on this time cutoff to ensure that we are able to reorient the systems such that we can reliably measure the spiral frequencies across time.
The precise time at which we are able to reorient each system varies across the simulations (see Table~\ref{tab:properties}); however, the bursty to steady transition demarcates the time when we are able to examine the time-evolution of the disk from snapshot to snapshot in the simulations.

We require the change in pattern speed between adjacent data points in time to be less than or equal to two times our minimum frequency resolution $\Delta\Omega \leq (2\cdot2\pi/3)\sim 4.2$~km/s/kpc to be considered members of the same spiral episode.
This threshold on $\Delta\Omega_p$ accommodates evolving pattern speeds.
Because we perform the WDFT on overlapping windows of time, it is reasonable to attribute incrementally changing pattern speeds to the same spiral amplitude.
This threshold is large enough to differentiate between spiral episodes of different frequencies but small enough to accommodate evolving frequencies.
Moreover, each data point can only be attributed to one spiral episode.

Additionally, we require that at least three consecutive data points meet the above criterion to be considered a spiral episode and included in our catalog (Table~\ref{tab:spiral-properties}).
We consider the lifetime of a spiral episode to be from the midpoints of the start and end time baseline $\mp 0.25$~Gyr.
Because the data points include overlapping windows of time, the minimum resolvable arm lifetime is thus $1.0$~Gyr.

We find that by implementing the above criteria, we are able to reliably identify spiral episodes and track the evolution of dominant amplitudes in a way that is consistent across all simulations.
The results of this procedure identifying spiral episodes are listed in Table~\ref{tab:spiral-properties}, and global summary statistics are listed in Table~\ref{tab:spiral-stats}.

We note in the case of m12b that we slightly modify the time that we begin identifying spiral episodes. 
We begin identification after 8.6~Gyr to avoid including features that we determined were not $m=2$ spiral arms through inspection of the corresponding spectrograms.
This is reflected by the grayed out region of m12b extending past the transition to steady star formation.

\section{Inverse Fourier Transform} \label{app:inverse}

To determine $a_{2\text{,max}}$ for a specified range of pattern speeds, we apply a band-pass filter in frequency space on the power spectrum. 
An inverse Fourier Transform is applied across the minimum and maximum frequency of the pattern $\pm \Delta\Omega$~km/s/kpc.
Thus, we filter out the power contribution of m=2 spiral amplitudes outside this range of $\Omega_p$.
To account for the windowing applied to the forward Fourier Transform, we employ an overlap-add method when performing the inverse Fourier Transform.

In this process, we perform the inverse Fourier Transform as
\begin{equation} \label{eq:9}
    c_m(R, t_j) = \frac{1}{S} \sum_{k=0}^{S-1} C_{m,k}(R) e^{2\pi ikj/S}
\end{equation}
where we set $C_{m,k}=0$ for values outside the frequency range of interest.
We generate these coefficients over each time baseline and add them to a running total.
The resulting total coefficients are then normalized to account for the windowing. 

Once we have the fully reconstructed amplitude over the desired frequency range, we determine the radius with max amplitude and corresponding time of $a_{2,max}$.
Due to evolving values of $R_{90}$, which are calculated at the midpoint of each time baseline, we only consider reconstructed Fourier coefficients when $R_{\text{peak}} < R_{90}$. 
This is to ensure that the $a_{2,max}$ identified is attributed to by the spiral amplitude and not extraneous power beyond the disk.
The values of $R_{\text{peak}}$ and time$_{\text{max}}$ reported in Table~\ref{tab:spiral-properties} corresponds to the radius and time of $a_{2,max}$.

\clearpage
\pagebreak

\section{Mass-weighted Identified m=2 Spirals} \label{app:mass_weight}

\begin{figure*}[h]
\begin{center}
\includegraphics[width=0.75\linewidth]{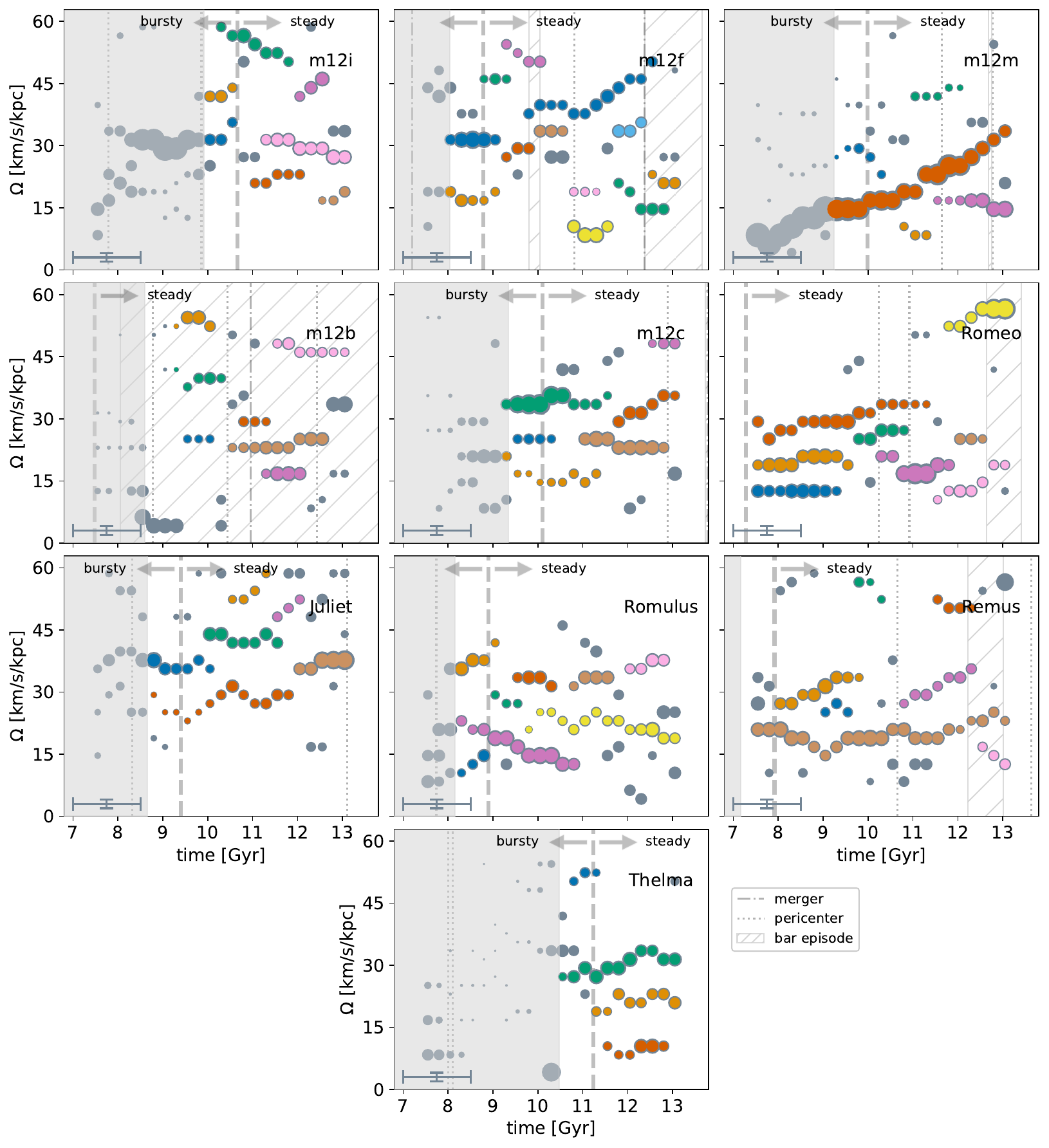}
\caption{Similar to Fig.~\ref{fig:all_sims_identified_spirals}, but applying mass normalization to the power spectrum.
Identified m=2 spiral episodes using the procedure described in Appendix~\ref{app:identify_arms} for all FIRE-2 galaxies analyzed.
Each color represents a distinct spiral episode. 
gray points represent m=2 spiral amplitudes present over the time baseline, but not identified as a member of a spiral episode.
\label{fig:all_sims_mass_weight}}
\end{center}
\end{figure*}

Fig.~\ref{fig:all_sims_identified_spirals} shows identified $m=2$ spiral episodes obtained by applying mass-weighting for the calculation of total power.  
Following the technique used in \cite{khachaturyants2022a,khachaturyants2022b}, we modify Eq.~\ref{eq:total_power} to be,
\begin{equation}
    \text{total power}(\Omega_k) = \frac{\sum M^2(R)P(R,\Omega_k)}{\sum M^2(R)}
\end{equation}
The advantage to mass-weighting the total power is the resulting emphasis on the relative strength of spirals across the disk rather than fractional power per annulus.
This method is useful for highlighting the perturbations that dominate the dynamics of the disk where most of mass exists.
In the body of this paper, we do not mass-weight, in keeping with the conventional approach \citep{sellwood&athanassoula1986,roskar2012}.
Each method has its advantages and can lead to a slight differences in the relative dominance of each spiral perturbation.
Overall we find that the strongest features are identified with both methods, but mass normalization identifies higher $\Omega_p$ spiral episodes that are located closer to the disk's central regions.
Fig.~\ref{fig:all_sims_identified_spirals} shows how mass-weighting impacts the results, and Table~\ref{tab:spiral-properties-mass-weight} (analogous to Table~\ref{tab:spiral-properties}) contains $m=2$ spiral properties obtained using the mass-weighted approach.
A deeper analysis of the dynamics of spirals in the disk is reserved in a future paper.

\startlongtable
\begin{deluxetable*}{cr>{\hspace{-7pt}- }lr>{\hspace{-7pt}- }lr>{\hspace{-7pt}- }lccccccccc}
\tablecaption{Properties of identified $m=2$ spiral episodes using the mass-weighted approach.\label{tab:spiral-properties-mass-weight}}
\tablehead{\vspace{-0.2cm}
  Sim & \multicolumn{2}{c}{Time} & \multicolumn{2}{c}{Radii} & \multicolumn{2}{c}{$\Omega_p$} & \colhead{$a_{2\text{,max}}$} & \colhead{$\text{time}_{\text{max}}$} & \colhead{$R_{\text{peak}}$} & \colhead{$R_{\text{ILR}}$} & \colhead{$R_{\text{CR}}$} & \colhead{$R_{\text{OLR}}$} & \colhead{$\Delta\text{time}$} & \colhead{$\Delta R$} & \colhead{Nearest} \vspace{-0.1cm} \\
   & & \colhead{} & \colhead{} & \colhead{} & \colhead{} & \colhead{} & \colhead{} & & & & & & & & \colhead{Resonance} \vspace{-0.5cm} \\
  & \multicolumn{2}{c}{[Gyr]} & \multicolumn{2}{c}{[kpc]} & \multicolumn{2}{c}{[km s$^{-1}$kpc$^{-1}$]}\hspace{-5pt} & & [Gyr] & [kpc] & [kpc] & [kpc] & [kpc] & [Gyr] & [kpc] & 
}
\startdata
\multirow{7}{*}{m12i} & 9.80 & 10.80 & 1.5 & 8.1 & 31.4 & 35.6 & 0.038 & 9.89 & 5.1 & 3.3 & 5.7 & 7.8 & 1.00 & 6.6 & CR \\
 & 9.80 & 10.80 & 1.2 & 7.8 & 41.9 & 44.0 & 0.037 & 10.04 & 5.1 & 2.7 & 4.8 & 6.3 & 1.00 & 6.6 & CR \\
 & 10.05 & 12.05 & 0.9 & 8.4 & 50.3 & 58.6 & 0.036 & 10.74 & 3.9 & 2.1 & 3.9 & 5.4 & 2.00 & 7.5 & CR \\
 & 10.80 & 12.30 & 3.3 & 9.0 & 20.9 & 23.0 & 0.055 & 11.65 & 9.0 & 4.5 & 7.5 & 10.2 & 1.50 & 5.7 & OLR \\
 & 11.80 & 12.80 & 1.2 & 9.1 & 41.9 & 46.1 & 0.016 & 12.20 & 4.5 & 2.7 & 4.8 & 6.3 & 1.00 & 7.9 & CR \\
 & 12.30 & 13.30 & 4.8 & 9.3 & 16.8 & 18.8 & 0.045 & 12.87 & 9.3 & 6.0 & 9.9 & 13.2 & 1.00 & 4.5 & CR \\
 & 11.05 & 13.30 & 2.1 & 8.6 & 27.2 & 31.4 & 0.043 & 11.19 & 3.9 & 3.6 & 6.3 & 8.4 & 2.25 & 6.5 & ILR \\ \hline
\multirow{12}{*}{m12f} & 7.80 & 9.30 & 2.1 & 9.4 & 31.4 & 31.4 & 0.049 & 8.58 & 3.0 & 3.3 & 5.7 & 7.8 & 1.50 & 7.3 & ILR \\
 & 7.80 & 9.30 & 2.1 & 9.2 & 16.8 & 18.8 & 0.074 & 8.27 & 5.7 & 5.4 & 9.0 & 12.0 & 1.50 & 7.1 & ILR \\
 & 8.55 & 9.55 & 2.1 & 9.0 & 46.1 & 46.1 & 0.030 & 8.16 & 2.1 & 2.4 & 4.2 & 5.7 & 1.00 & 6.9 & ILR \\
 & 9.05 & 10.05 & 2.1 & 9.2 & 27.2 & 29.3 & 0.051 & 8.66 & 3.6 & 3.6 & 6.0 & 8.1 & 1.00 & 7.1 & ILR \\
 & 9.05 & 10.30 & 2.1 & 9.9 & 50.3 & 54.5 & 0.038 & 9.30 & 3.3 & 2.4 & 3.9 & 5.4 & 1.25 & 7.8 & CR \\
 & 9.80 & 10.80 & 2.1 & 11.2 & 33.5 & 33.5 & 0.033 & 10.04 & 6.3 & 3.3 & 5.7 & 7.8 & 1.00 & 9.1 & CR \\
 & 10.55 & 11.55 & 3.6 & 12.0 & 18.8 & 18.8 & 0.059 & 10.07 & 12.0 & 5.1 & 8.7 & 11.7 & 1.00 & 8.4 & OLR \\
 & 10.55 & 11.80 & 5.1 & 12.0 & 8.4 & 10.5 & 0.190 & 11.11 & 12.0 & 9.0 & - & - & 1.25 & 6.9 & ILR \\
 & 11.55 & 12.55 & 2.1 & 11.7 & 33.5 & 35.6 & 0.033 & 11.06 & 3.6 & 3.3 & 5.7 & 7.8 & 1.00 & 9.6 & ILR \\
 & 9.55 & 12.80 & 2.1 & 11.2 & 37.7 & 50.3 & 0.069 & 9.97 & 2.1 & 3.0 & 5.1 & 6.9 & 3.25 & 9.1 & ILR \\
 & 12.30 & 13.30 & 3.0 & 12.7 & 20.9 & 23.0 & 0.050 & 12.87 & 5.4 & 5.1 & 8.7 & 11.7 & 1.00 & 9.7 & ILR \\
 & 11.55 & 13.05 & 4.2 & 12.9 & 14.7 & 20.9 & 0.099 & 12.45 & 7.8 & 6.6 & 11.1 & - & 1.50 & 8.7 & ILR \\ \hline
\multirow{5}{*}{m12m} & 9.05 & 10.55 & 1.5 & 12.2 & 23.0 & 29.3 & 0.060 & 9.94 & 9.3 & 3.9 & 7.5 & 10.5 & 1.50 & 10.7 & OLR \\
 & 10.55 & 11.55 & 4.5 & 12.3 & 8.4 & 10.5 & 0.078 & 10.72 & 12.0 & 11.4 & - & - & 1.00 & 7.8 & ILR \\
 & 10.80 & 12.30 & 1.5 & 12.3 & 41.9 & 44.0 & 0.028 & 10.32 & 7.5 & 2.7 & 5.1 & 7.2 & 1.50 & 10.8 & OLR \\
 & 9.05 & 13.30 & 1.5 & 12.3 & 14.7 & 33.5 & 0.164 & 9.20 & 3.0 & 4.8 & 9.0 & 12.3 & 4.25 & 10.8 & ILR \\
 & 11.30 & 13.30 & 3.9 & 12.3 & 14.7 & 16.8 & 0.064 & 12.38 & 12.3 & 6.6 & 11.7 & - & 2.00 & 8.4 & CR \\ \hline
\multirow{7}{*}{m12b} & 9.30 & 10.30 & 2.1 & 6.9 & 25.1 & 25.1 & 0.054 & 10.29 & 5.7 & 4.5 & 7.5 & 9.9 & 1.00 & 4.8 & ILR \\
 & 9.05 & 10.30 & 2.1 & 6.1 & 52.4 & 54.5 & 0.046 & 9.07 & 3.6 & 2.7 & 4.2 & 5.7 & 1.25 & 4.0 & CR \\
 & 9.05 & 10.55 & 2.1 & 6.6 & 37.7 & 41.9 & 0.067 & 8.94 & 6.6 & 3.3 & 5.4 & 6.9 & 1.50 & 4.5 & OLR \\
 & 10.55 & 11.55 & 3.0 & 10.5 & 29.3 & 29.3 & 0.064 & 10.91 & 10.5 & 4.2 & 6.9 & 9.3 & 1.00 & 7.5 & OLR \\
 & 11.05 & 12.30 & 2.1 & 10.2 & 16.8 & 16.8 & 0.093 & 11.60 & 7.8 & 6.3 & 10.5 & 14.1 & 1.25 & 8.1 & ILR \\
 & 10.30 & 12.80 & 2.7 & 10.5 & 23.0 & 25.1 & 0.104 & 10.96 & 7.5 & 4.8 & 8.1 & 11.1 & 2.50 & 7.8 & CR \\
 & 11.30 & 13.30 & 2.1 & 7.2 & 46.1 & 48.2 & 0.028 & 11.65 & 3.3 & 3.0 & 5.1 & 6.6 & 2.00 & 5.1 & ILR \\ \hline
\multirow{4}{*}{m12c} & 9.30 & 10.55 & 1.5 & 9.0 & 25.1 & 25.1 & 0.046 & 9.91 & 7.8 & 3.6 & 6.3 & 8.7 & 1.25 & 7.5 & OLR \\
 & 9.05 & 11.55 & 3.6 & 9.0 & 14.7 & 20.9 & 0.079 & 11.74 & 9.0 & 5.4 & 9.0 & 12.3 & 2.50 & 5.4 & CR \\
 & 9.05 & 11.80 & 1.5 & 9.0 & 33.5 & 35.6 & 0.078 & 10.04 & 1.5 & 2.7 & 5.1 & 6.9 & 2.75 & 7.5 & ILR \\
 & 11.55 & 13.30 & 1.5 & 9.1 & 29.3 & 35.6 & 0.039 & 11.36 & 7.5 & 3.0 & 5.4 & 7.5 & 1.75 & 7.6 & OLR \\
 m12c & 12.30 & 13.30 & 1.5 & 9.9 & 48.2 & 48.2 & 0.013 & 12.96 & 4.8 & 2.1 & 4.2 & 5.7 & 1.00 & 8.4 & CR \\
 (\textit{cont.}) & 10.80 & 13.05 & 2.1 & 9.6 & 23.0 & 25.1 & 0.075 & 11.50 & 9.6 & 4.2 & 7.2 & 9.6 & 2.25 & 7.5 & OLR \\ \hline
\multirow{8}{*}{Romeo} & 7.30 & 9.55 & 5.1 & 11.4 & 12.6 & 12.6 & 0.077 & 8.70 & 11.4 & 6.6 & 11.4 & - & 2.25 & 6.3 & CR \\
 & 7.30 & 9.80 & 3.0 & 11.1 & 18.8 & 20.9 & 0.091 & 7.95 & 11.1 & 4.8 & 8.1 & 11.1 & 2.50 & 8.1 & OLR \\
 & 9.55 & 11.05 & 2.7 & 12.1 & 25.1 & 27.2 & 0.032 & 10.65 & 10.5 & 3.6 & 6.3 & 8.7 & 1.50 & 9.4 & OLR \\
 & 7.30 & 11.55 & 2.4 & 11.1 & 25.1 & 33.5 & 0.064 & 8.60 & 6.9 & 3.3 & 6.0 & 8.1 & 4.25 & 8.7 & CR \\
 & 10.05 & 12.05 & 4.2 & 12.6 & 16.8 & 20.9 & 0.054 & 10.68 & 12.6 & 5.1 & 8.7 & 11.7 & 2.00 & 8.4 & OLR \\
 & 11.80 & 12.80 & 2.7 & 12.8 & 25.1 & 25.1 & 0.022 & 12.26 & 9.0 & 3.9 & 6.9 & 9.6 & 1.00 & 10.1 & OLR \\
 & 11.30 & 13.30 & 4.8 & 12.9 & 10.5 & 18.8 & 0.054 & 12.45 & 8.1 & 6.9 & 12.0 & - & 2.00 & 8.1 & ILR \\
 & 11.55 & 13.30 & 1.5 & 6.9 & 52.4 & 56.5 & 0.023 & 12.95 & 1.5 & 2.1 & 3.6 & 5.1 & 1.75 & 5.4 & ILR \\ \hline
\multirow{6}{*}{Juliet} & 8.55 & 10.30 & 0.3 & 7.2 & 35.6 & 37.7 & 0.040 & 8.91 & 1.8 & 3.0 & 5.1 & 6.9 & 1.75 & 6.9 & ILR \\
 & 10.30 & 11.55 & 3.6 & 3.9 & 52.4 & 58.6 & 0.016 & 10.98 & 3.6 & 2.1 & 3.9 & 5.4 & 1.25 & 0.3 & CR \\
 & 9.80 & 11.80 & 1.5 & 7.1 & 41.9 & 44.0 & 0.024 & 9.84 & 2.1 & 2.4 & 4.5 & 6.3 & 2.00 & 5.6 & ILR \\
 & 8.55 & 12.05 & 3.0 & 7.2 & 23.0 & 31.4 & 0.027 & 11.43 & 6.3 & 3.9 & 6.6 & 9.0 & 3.50 & 4.2 & CR \\
 & 11.30 & 12.30 & 3.3 & 4.5 & 48.2 & 52.4 & 0.012 & 11.84 & 3.9 & 2.4 & 4.2 & 5.7 & 1.00 & 1.2 & CR \\
 & 11.80 & 13.30 & 0.3 & 6.9 & 35.6 & 37.7 & 0.014 & 12.01 & 6.3 & 3.0 & 5.1 & 7.2 & 1.50 & 6.6 & OLR \\ \hline
\multirow{8}{*}{Romulus} & 8.05 & 9.05 & 4.5 & 11.1 & 10.5 & 14.7 & 0.072 & 8.61 & 8.1 & 7.5 & 12.3 & - & 1.00 & 6.6 & ILR \\
 & 8.05 & 9.30 & 2.1 & 9.9 & 35.6 & 41.9 & 0.055 & 8.48 & 3.0 & 3.0 & 5.4 & 7.5 & 1.25 & 7.8 & ILR \\
 & 8.80 & 9.80 & 3.0 & 10.8 & 27.2 & 29.3 & 0.044 & 8.32 & 9.3 & 3.9 & 6.9 & 9.3 & 1.00 & 7.8 & OLR \\
 & 9.30 & 10.55 & 2.7 & 11.4 & 31.4 & 33.5 & 0.031 & 9.93 & 6.3 & 3.6 & 6.0 & 8.4 & 1.25 & 8.7 & CR \\
 & 8.05 & 11.05 & 0.3 & 11.3 & 12.6 & 23.0 & 0.106 & 8.43 & 8.1 & 6.0 & 9.9 & 13.5 & 3.00 & 11.0 & CR \\
 & 10.55 & 11.80 & 3.0 & 12.3 & 31.4 & 33.5 & 0.034 & 10.87 & 9.6 & 3.6 & 6.0 & 8.4 & 1.25 & 9.3 & OLR \\
 & 11.80 & 13.05 & 2.4 & 12.8 & 35.6 & 37.7 & 0.020 & 11.31 & 3.0 & 3.3 & 5.7 & 7.8 & 1.25 & 10.4 & ILR \\
 & 9.55 & 13.30 & 4.5 & 14.1 & 18.8 & 25.1 & 0.067 & 12.58 & 13.8 & 5.1 & 8.7 & 11.7 & 3.75 & 9.6 & OLR \\ \hline
\multirow{7}{*}{Remus} & 8.80 & 9.80 & 2.1 & 10.0 & 25.1 & 27.2 & 0.042 & 8.37 & 9.0 & 3.3 & 5.7 & 7.8 & 1.00 & 7.9 & OLR \\
 & 7.80 & 10.05 & 1.8 & 10.0 & 27.2 & 33.5 & 0.039 & 8.40 & 8.1 & 3.0 & 5.1 & 6.9 & 2.25 & 8.2 & OLR \\
 & 9.55 & 10.55 & 2.4 & 4.5 & 52.4 & 56.5 & 0.018 & 9.86 & 3.3 & 1.8 & 3.0 & 4.5 & 1.00 & 2.1 & CR \\
 & 11.30 & 12.55 & 1.8 & 3.9 & 50.3 & 52.4 & 0.010 & 11.55 & 1.8 & 2.1 & 3.6 & 4.8 & 1.25 & 2.1 & ILR \\
 & 10.55 & 12.55 & 1.8 & 10.9 & 27.2 & 35.6 & 0.027 & 10.28 & 8.1 & 3.0 & 5.1 & 6.9 & 2.00 & 9.1 & OLR \\
 & 7.30 & 13.30 & 2.7 & 9.9 & 14.7 & 25.1 & 0.071 & 8.30 & 9.9 & 4.2 & 7.2 & 9.9 & 6.00 & 7.2 & OLR \\
 & 12.30 & 13.30 & 3.9 & 12.0 & 12.6 & 16.8 & 0.029 & 12.65 & 7.5 & 5.4 & 9.3 & 12.6 & 1.00 & 8.1 & CR \\ \hline
\multirow{4}{*}{Thelma} & 10.55 & 11.55 & 2.1 & 9.3 & 50.3 & 52.4 & 0.021 & 11.02 & 4.5 & 2.1 & 3.6 & 5.1 & 1.00 & 7.2 & OLR \\
 & 11.05 & 13.30 & 1.8 & 11.0 & 18.8 & 23.0 & 0.040 & 11.29 & 10.2 & 4.5 & 8.1 & 10.8 & 2.25 & 9.2 & OLR \\
 & 10.30 & 13.30 & 1.5 & 11.1 & 27.2 & 33.5 & 0.039 & 11.16 & 7.5 & 3.3 & 6.0 & 8.4 & 3.00 & 9.6 & OLR \\
 & 11.30 & 13.05 & 4.5 & 11.4 & 8.4 & 10.5 & 0.081 & 12.60 & 11.4 & 8.4 & 13.8 & - & 1.75 & 6.9 & CR 
\enddata
\vspace{-0.15cm}
\tablecomments{Columns: (1) Name of the galaxy. 
(2) Time: start and end time of the spiral episode, listed at the midpoint of the start and end time baselines $\mp 0.25$~Gyr. 
(3) Radii: Radial range of the spiral episode. 
(4) $\Omega_p$: Pattern speed range of the spiral episode. 
(5) $a_{2,\text{max}}$: Maximum $a_2$ amplitude of the frequency filtered inverse Fourier transform (see \S\ref{app:identify_arms}). 
(6) Time$_{\text{max}}$: Time of $a_{2\text{,max}}$. 
(7) $R_{\text{peak}}$: Radius at which we identify $a_{2\text{,max}}$. 
(8) $R_{\text{ILR}}$: Radius of Inner Linblad resonance at Time$_{\text{max}}$. 
(9) $R_{\text{CR}}$: Corotation radius at Time$_{\text{max}}$. 
(10) $R_{\text{OLR}}$: Radius of Outer Linblad resonance Time$_{\text{max}}$. 
Blank entries indicate that the resonance radius is beyond the edge of the disk. 
(11) $\Delta\text{time}$: Duration of the spiral episode, calculated from column 2. 
(12) $\Delta R$: Radial extent of the spiral episode. 
(13) Resonance point nearest to $R_{\text{peak}}$.}
\end{deluxetable*}

\vspace{-0.5cm}
\section{WDFT of Stars and Cold Dense Gas} \label{app:stars_cdgas}

\begin{figure*}[h]
\begin{center}
\includegraphics[width=0.75\linewidth]{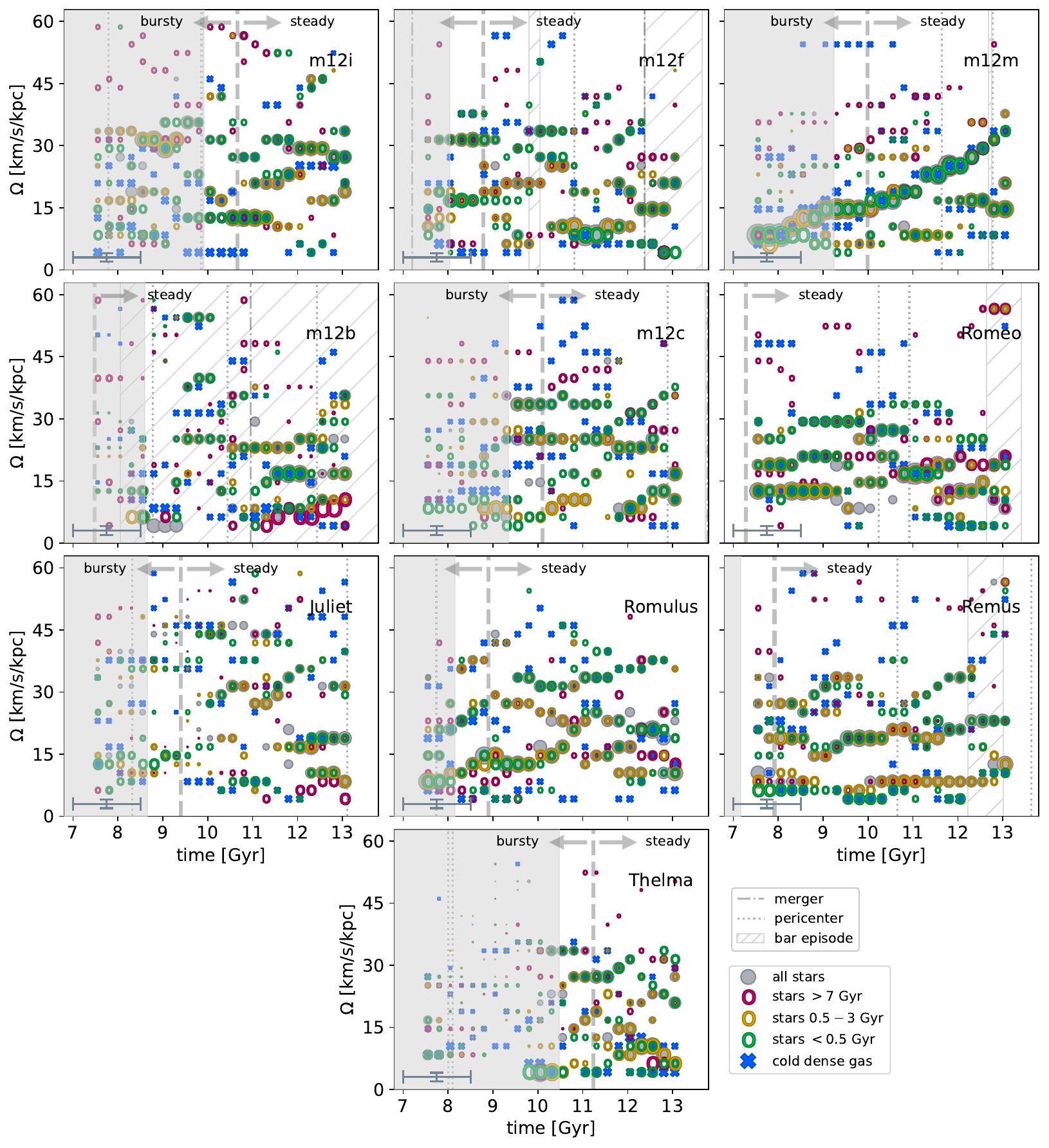}
\caption{Expanded version of the composite panels in Fig.~\ref{fig:bubble_stars_gas_conjoint_m12f_m12m}.
Frequency evolution of the three most dominant m=2 amplitudes in stars with varying stellar ages and star forming gas ($T<10^4$ K and amu/cm$^3>10$) for all FIRE-2 galaxies analyzed.
Stars are divided into three different groups of stellar ages: $<0.5$~Gyr (green), $0.5-3$~Gyr (yellow), and $>7$~Gyr (magenta). 
The cold dense gas is shown in the filled blue crosses. 
The gray points show all stars and are the same as in Fig.~\ref{fig:all_sims_identified_spirals}.
Points are sized proportional to their amplitudes in each time baseline.
The thick gray vertical dashed line indicates the transition from bursty to steady star formation rate.
The grayed out region shows the time period that we do not analyze the results for the galaxy.
We indicate the time of merger events (dash-dot line), pericentric passages (dotted line), and bar episodes (hatched rectangle).
\label{fig:all_sims_bubble_stars_cdgas}}
\end{center}
\end{figure*}

We show the evolution of $m=2$ frequencies in stars with varying stellar ages and cold dense gas in Fig.~\ref{fig:all_sims_bubble_stars_cdgas}.
In many cases, all populations of stars and gas follow the same evolution of $\Omega_p$.
Notably, stars $>7$~Gyr in m12b largely deviate in $\Omega_p$ from other populations, more so than in any other galaxy.
The gas overall traces the stars remarkably well given the distribution of cold dense gas in FIRE-2 is somewhat clumpy, and thus the WDFT may not be as effective at tracking the cohesive evolution of $\Omega_p$.
The largest discrepancies between stars and cold dense gas are in m12c, Juliet, and Thelma, all three of which have relatively kinematically hotter disks.

\end{document}